\documentclass[twocolumn,prl,aps,superscriptaddress,tightenlines]{revtex4}
\usepackage{amsfonts}
\usepackage{amssymb}
\usepackage{epsfig}
\usepackage{amsbsy}
\usepackage{amsmath}
\usepackage{graphicx}
\usepackage{CJK}
\usepackage{graphicx}
\usepackage{dcolumn}
\usepackage{bm}

\begin{document}

\title{Topological Phenomena in Classical Optical Networks}
\author{T. Shi$^{1}$, H. J. Kimble$^{1,2,3}$, and J. I. Cirac}
\affiliation{Max-Planck-Institut f\"{u}r Quantenoptik, Hans-Kopfermann-Str. 1, 85748
Garching, Germany \\
$^{2}$Norman Bridge Laboratory of Physics 12-33 \\
$^{3}$Institute for Quantum Information and Matter, California Institute of
Technology, Pasadena, CA 91125, USA}

\begin{abstract}
We propose a scheme to realize a topological insulator with optical-passive
elements, and analyze the effects of Kerr-nonlinearities in its topological
behavior. In the linear regime, our design gives rise to an optical spectrum
with topological features and where the bandwidths and bandgaps are
dramatically broadened. The resulting edge modes cover a very wide frequency
range. We relate this behavior to the fact that the effective Hamiltonian
describing the system's amplitudes is long-range. We also develop a method
to analyze the scheme in the presence of a Kerr medium. We assess robustness
and stability of the topological features, and predict the presence of
chiral squeezed fluctuations at the edges in some parameter regimes.
\end{abstract}

\maketitle

\textit{Introduction.}---The discovery of topological insulators (TI), as
well as Quantum Spin Hall (QSH) insulators \cite%
{TKNN,TI-Haldane,TI-Hasan,TI-Qi,QSHT,HgTeQW,QSH,SDsurface,QAHE} have opened
up a wide range of scientific and technological questions. Their spectra
(dispersion relation) feature a set of bands, connected by chiral edge modes
that reflect the topological nature of the material. These modes are robust
against perturbations whose energy does not exceed the corresponding bandgap
and that do not break the time-reversal (TR) symmetry, in the case of the
QSH \cite{Z2,spinpump}. Electronic interactions give rise to a wide range of
phenomena. Although the edge modes persist, their properties are
qualitatively modified \cite{Edgemodes}. In addition, they can give rise to
other exotic phenomena, like the fractionalization of charges, or the
appearance of excitations with fractional statistics \cite%
{Abelian,Nonabelian}.

Recent proposals to generate TI and QSH with (classical) light have also
attracted a lot of attention \cite%
{imaging,FTI,coldatom,trappedion,TIph-Haldane,TIph-Wang,emag,FTI2,DGF,TOM,pFQHE,opticaldelay,Agauge,BM}%
. In fact, the first experimental observations \cite{imaging,FTI} of
topological features in optical systems have been recently reported, and a
several schemes exhibiting intriguing features have been proposed \cite%
{coldatom,trappedion,TIph-Haldane,TIph-Wang,emag,FTI2,DGF,TOM,pFQHE}. There
exist different setups where one can realize the optical analog of QSH, and
observe similar features. In the context of coupled resonator arrays, one
can use either differential optical paths in waveguides \cite{opticaldelay}
or an optical active (Faraday rotator) element \cite{Agauge}. Despite their
success, in the first case it would be desirable to enlarge the bandgaps in
the spectrum, which is limited by the small coupling of the local modes in
the (high finesse) resonators \cite{opticaldelay}, in order to gain
robustness. In the second, photon absorption in the active media also limits
the operationality of the scheme. In other schemes, like the one based on
bianisotropic metacrystals \cite{BM}, the realization of long-lived edge
modes in a broader frequency range is challenged by the weak bianisotropy in
metamaterials \cite{bianisotropy1,bianisotropy2}. The effects of
interactions in those optical models have been scarcely investigated \cite%
{TOM,NL1,NL2}. In particular, the fate of the edge modes, their stability,
or the appearance of quantum features has not been analyzed so far.

\begin{figure}[tbp]
\includegraphics[bb=95 309 545 710, width=8 cm, clip]{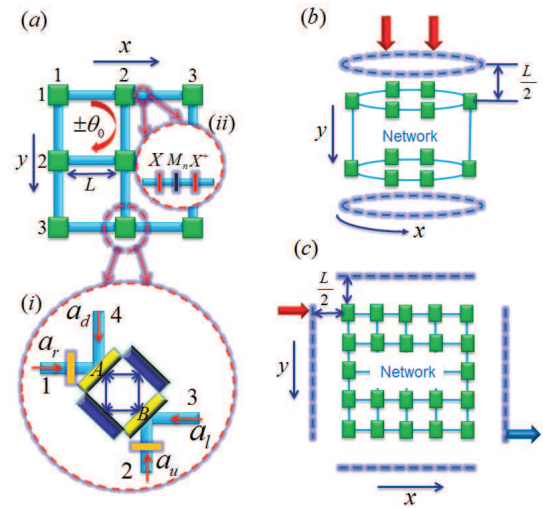}
\caption{(Color Online) Schematic of our optical network: (a) Two insets
show the construction of optical elements in nodes and fibers, respectively.
Here, the light acquires a polarization-dependent phase $\protect\theta_0$
by propagating clockwise around each plaquette (the red arrow). (b)-(c)
Perfect and partially transmissive mirrors are put along the boundaries of
the cylindrical and planar networks. Here, the partially transmissive
mirrors are placed along the top boundary in the cylindrical network, and at
the top-left and bottom-right corners in the planar network. The driving
light (red arrows) is applied to generate the bulk and edge excitations, and
the transmitted light is denoted by the blue arrow. Detailed descriptions of
the various elements are given in Ref. \protect\cite{SM}.}
\label{schematic}
\end{figure}

In this letter we propose and analyze a scheme to realize the optical
version of the QSH insulator, and investigate the effects produced by
Kerr-nonlinearities, which play the role by photon-photon interactions. Our
scheme uses beam splitters, lambda plates, and birrefringent materials, that
are optically passive and thus circumvent the problem of photon absorption.
In contrast to previous schemes, the effective Hamiltonian description of
our setup features long-range hopping behavior, which dramatically broadens
the spectral bands and bandgaps, thus adding robustness to perturbations. In
fact, the edge modes cover a large frequency range. They can be excited with
optical driving, and detected through the measurement of phase shifts and
the transmission spectrum. In the presence of a Kerr medium, the chiral edge
modes survive, can be externally excited, and are stable in some parameter
regimes which we identify. By additionally probing with a different light
frequency, edge (Bogoliubov) excitations can be excited. As an interplay
between the Kerr-nonlinearity and the topology, they are also chiral and
possess quantum features, namely they are squeezed. In order to obtain these
results, we develop a scattering matrix approach, and show how to solve it
with numerical methods.

\textit{Model setup.}---The architecture of the non-trivial topology is
designed in the 2D network with size $N_{x}\times N_{y}$ by the linear
optics elements, i.e., optical fibers, birefringent elements, beamsplitters
and perfect mirrors. As shown in Fig. \ref{schematic}, at each node of a
square lattice, two beamsplitters with reflectance $R_{\mathrm{bs}}=(\sqrt{2}%
-1)^{2}\sim 0.17$ and two perfect mirrors form a \textquotedblleft bad
cavity\textquotedblright\ to change the propagation direction of the light,
and nearest neighbor nodes are connected by one optical fiber with length $L$%
. The inset (i) in Fig. \ref{schematic}a shows that the fibers $1$ and $4$
are connected to the beamsplitter $A$, while the fibers $2$ and $3$ are
connected to the beamsplitter $B$. Two birefringent elements next to the
node in the fibers $1$ and $2$ cancel the sign change of the vertical
polarization for the light propagating through the node (Sec. I in \cite{SM}%
).

The inset (ii) in Fig. \ref{schematic}a shows that in each horizontal fiber,
three birefringent elements are applied to generate an artificial
\textquotedblleft magnetic\textquotedblright\ field for the light
propagating in the network. From left to right, the three birefringent
elements are described by the Jones matrices $X=e^{i\pi \sigma _{x}/4}$, $%
M_{n}=e^{in\theta _{0}\sigma _{z}}$, and $X^{\dagger }$ defined in the
linear polarization basis $(H,V)$, where $\sigma _{x,y,z}$ are Pauli
matrices, and the phase shift $n\theta _{0}$ depends on the row index $n$.
The interaction of photons is induced by a Kerr nonlinearity in the fiber,
where the light propagating in the fiber acquires an additional phase
proportional to the light intensity.

Different boundary conditions give rise to a rich variety of geometries for
the network.\ As shown in Figs. \ref{schematic}b and \ref{schematic}c, for
the cylindrical and planar geometries, perfect mirrors are placed along the
boundaries to form the closed network, where the distance between boundary
mirrors and boundary nodes is $L/2$. In the open network with partially
transmissive mirrors on the boundary, non-trivial topological states can be
generated and detected by external fields injected through the boundary
mirrors.

\textit{Linear Regime.}--- The birefringent elements cause the light to
acquire a phase matrix $\theta _{0}\sigma _{y}$ by propagating around each
plaquette. Hence, the system is TR-invariant, namely, circular polarizations
($\sigma _{\pm }$) experience oppositely directed \textquotedblleft
magnetic\textquotedblright\ fields with fluxes $\pm \theta _{0}$,
respectively, which is the origin of nontrivial topology in the network of
linear optical elements.

\begin{figure}[tbp]
\includegraphics[bb=73 282 517 700, width=7.5 cm, clip]{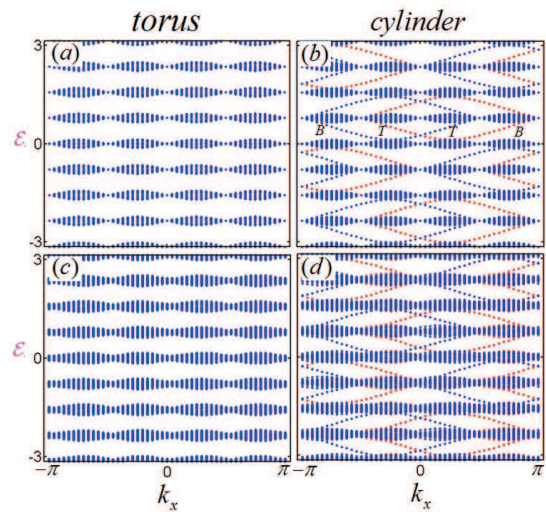}
\caption{(Color Online) The energy spectra on the torus and cylinder, where $%
\protect\theta_{0}=\protect\pi/2$, the network size is $48 \times 48$, and $%
L $ is taken as unit. The blue and red colors denote $\protect\sigma_{+}$
and $\protect\sigma_{-}$ polarizations, respectively. (a)-(b) The energy
spectrum $\mathcal{E}$ on the torus (a) and the cylinder (b) without phase
randomness and zero losses of linear elements, where $B$ and $T$ denote the
bottom and top boundaries, respectively. (c)-(d) The energy spectrum Re$%
\mathcal{E}$ on the torus (c) and the cylinder (d) with phase randomness and
nonzero losses of linear elements.}
\label{linearspectrum}
\end{figure}

The photonic spectra $\mathcal{E}$ of linear network with different
geometries display non-trivial topologies of the light, which are determined
by the scattering equation $S_{0}\Psi _{nm}=e^{-i\mathcal{E}L}\tilde{\Psi}%
_{nm}$ (Sec. IIA and III in \cite{SM}). Here, the free $S$-matrix $S_{0}$
connects the right-, up-, left-, and down- moving input fields $\Psi
_{nm}=(a_{r,nm},a_{u,nm},a_{l,nm},a_{d,nm})^{T}$ at the node $(n,m)$ with
the input amplitudes $\tilde{\Psi}%
_{nm}=(a_{r,nm+1},a_{u,n-1,m},a_{l,nm-1},a_{d,n+1m})^{T}$ at four nearest
neighbor nodes. The eigenmode has the definite $\sigma _{+}$ or $\sigma _{-}$
polarization due to the block structure of $S_{0}$ in the basis of circular
polarizations. Figures \ref{linearspectrum}a-b show the spectra of networks
on the torus and the cylinder in one period $\mathcal{E}\in \omega
_{c}+(-\pi /L,\pi /L]$ around a large central frequency $\omega _{c}=2\pi
N_{c}/L$ ($N_{c}$ is an integer), where $\theta _{0}=\pi /2$, and $k_{x}$ is
the quasi-momentum of the eigenmode $%
a_{(r,u,l,d),nm}=a_{(r,u,l,d),n}e^{ik_{x}m}/\sqrt{N_{x}}$.

As shown in Fig. \ref{linearspectrum}a, for the torus geometry, the photonic
bands with broad gaps span across the whole spectral period $2\pi /L$. In
contrast with narrow-band schemes \cite{opticaldelay,Agauge}, the wide
spectrum results from the large hopping strength (comparable with $2\pi /L$)
between nodes (i.e., cavities) beyond nearest neighbors in the
\textquotedblleft bad cavity\textquotedblright\ regime $R_{\mathrm{bs}}\sim
0.17$. This long-range hopping behavior is characterized by the spatially
nonlocal Hamiltonian $H_{\mathrm{eff}}=i$ln$S_{0}/L$ rather than the
Hofstadter (tight binding) model \cite{HM}.

As a consequence of the bands with the non-trivial $Z_{2}$ topological index
\cite{Z2}, the helical edge modes arise in the broad topological band gaps.
In Fig. \ref{linearspectrum}b, for the cylindrical geometry, the spectrum
displays the four edge modes between band gaps, where the chiralities of two
edge modes on each boundary are locked to the $\sigma _{\pm }$
polarizations. The topologically protected helical edge modes are immune to
any local perturbation, randomness, and losses, as long as the band gap
remains open. Figures \ref{linearspectrum}c-d show that for the random phase
fluctuation $\delta _{p}\in \lbrack -0.2,0.2]$ around $\theta _{0}=\pi /2$
and the optical losses (10\%) of each element, the band gaps in the energy
spectrum Re$\mathcal{E}$ are still opened on the torus, and the helical edge
modes survive on the cylinder with a long life-time $\tau =1/$Im$\mathcal{E}%
\sim 13L$ (the speed $c$ of light is taken as the unit).

\begin{figure}[tbp]
\includegraphics[bb=64 330 532 757, width=8 cm, clip]{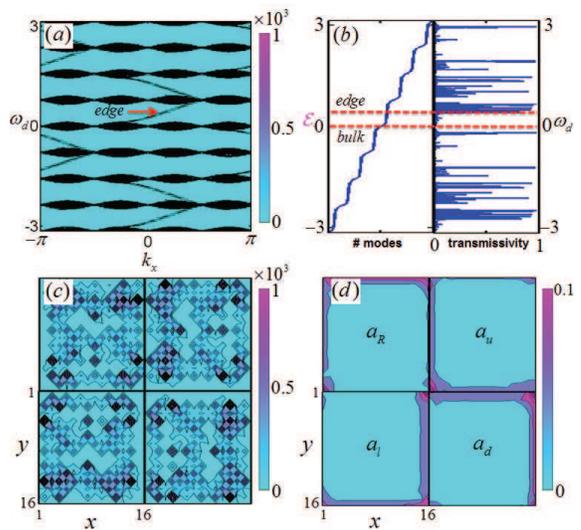}
\caption{(Color Online) Detection of topological properties, where $\protect%
\theta_{0}=\protect\pi/2$, $r_{\mathrm{BM}}=0.9$, and $L$ is taken as unit.
(a) For the cylindrical geometry, the contourplot of $d\Delta /d\protect%
\omega _{\mathrm{d}}$ shows the eigen-spectrum for the network of size $48
\times 48$; (b) For the open plane of size $16 \times 16$, the
eigen-spectrum for the closed network, and the transmission spectrum;
(c)-(d): The light intensities $\left\vert a_{(r,u,l,d)}/A_{\mathrm{in}%
}^{(0)}\right\vert ^{2}$ of (c) bulk and (d) edge modes in the network under
$\protect\sigma_{+} $-polarized driving light, where the first (second) row
shows the right- (left-) and up- (down-) moving internal fields.}
\label{lineardetection}
\end{figure}

In order to generate and probe the edge and bulk modes, input light $A_{%
\mathrm{in},m}^{(0)}e^{-i\omega _{\mathrm{d}}t}$ with amplitude $A_{\mathrm{%
in},m}^{(0)}=A_{\mathrm{in}}^{(0)}e^{ik_{x}m}/\sqrt{N_{x}}$ and frequency $%
\omega _{\mathrm{d}}$ is applied to drive the cylindrical network (Fig. \ref%
{schematic}b) through each of the transmissive top-boundary mirrors with the
reflection coefficient $r_{\mathrm{BM}}$. Significantly, the relative phase
shift $\Delta -(\pi +2\arctan r_{\mathrm{BM}})\in (-\pi ,\pi ]$ (Sec. IIIB
in \cite{SM}) of reflected light from the network jumps from $-\pi $ to $\pi
$, when $\omega _{\mathrm{d}}$ sweeps across a resonant frequency $\mathcal{E%
}$ of the closed network. Thus, the measurement of this phase shift reveals
the spectrum. As shown in Fig. \ref{lineardetection}a, for $\sigma _{+}$
polarized driving light, the peaks of $d\Delta /d\omega _{\mathrm{d}}$ show
the band structure and the chiral edge mode on the top boundary. Since the
bottom edge mode decays exponentially along the $y$-direction, its
decoupling to the driving light makes the bottom edge mode invisible in Fig. %
\ref{lineardetection}a, which isolates a single $\sigma _{+}$ polarized
chiral edge mode on the top boundary.

For the planar network, the circularly polarized driving light with
amplitude $A_{\mathrm{in}}^{(0)}$ and frequency $\omega _{\mathrm{d}}$ is
injected through the transmissive mirror at the upper-left corner (Fig. \ref%
{schematic}c). As shown in Fig. \ref{lineardetection}b, the transmission
spectrum of output light through the mirror at the bottom-right corner
displays the energy spectrum of the planar network. For $\omega _{\mathrm{d}%
}\sim 0.03/L$ ($0.37/L$) resonant with the bulk (edge) mode, as shown in
Fig. \ref{lineardetection}c (d), the light intensities $\left\vert
a_{(r,u,l,d),nm}/A_{\mathrm{in}}^{(0)}\right\vert ^{2}$ display that the
light propagates in the bulk and along the boundary, respectively.

\textit{Non-linear regime.}--- The nonlinear Kerr medium with the
self-focusing effect \cite{textbook,selffocusing} on the right hand side of
three birefringent elements generates an attractive interaction $\chi <0$ of
photons with the same polarization. Here, we consider separately $\sigma
_{\pm }$ polarizations found for the eigenmodes of the linear system, and
thereby avoid the complexity of nonlinear birefringence for $\sigma _{\pm }$
polarizations propagating simultaneouly in the fiber links with $\chi $ \cite%
{textbook,NOP}. In the horizontal link between nodes $(n,m)$ and $(n,m-1)$,
light propagation in the Kerr medium is described\ by the right- and left-
moving fields $\phi _{r,nm}(x,t)$ and $\phi _{l,nm-1}(x,t)$, which obey a
set of nonlinear motion equations \cite%
{instability1,instability2,instability3}. Similarly, the up- and down-
moving fields $\phi _{u,nm}(x,t)$ and $\phi _{d,n+1m}(x,t)$ describe the
light propagation in the vertical fiber connecting nodes $(n,m)$ and $%
(n+1,m) $. Here, $x$ is the coordinate along the fiber.

Under the circularly polarized driving field $A_{\mathrm{in}%
,m}^{(0)}e^{-i\omega _{\mathrm{d}}t}$ with amplitude $A_{\mathrm{in}%
,m}^{(0)}=A_{\mathrm{in}}^{(0)}e^{ik_{x}m}/\sqrt{N_{x}}$ through the top
boundary of cylindrical network, the fields $\phi _{s,nm}^{(0)}(x,t)=\phi
_{s,nm}^{(0)}(x)e^{-i\omega _{\mathrm{d}}t}$ ($s=r,u,l,d$) in the steady
state are plane waves $\phi
_{r(u),nm}^{(0)}(x)=a_{r(u),nm}e^{ik_{r(u)}(x-L)} $, $\phi
_{l,nm}^{(0)}(x)=e^{-in\sigma \theta _{0}}a_{l,nm}e^{-ik_{l}x}$, and $\phi
_{d,nm}^{(0)}(x)=a_{d,nm}e^{-ik_{d}x}$, where $\omega _{\mathrm{d}}-k_{s}$
is determined by the elements of $4\times 4$ diagonal matrices $\mathcal{N}%
_{nm}$ proportional to the light intensities $\left\vert
a_{r,u,l,d}\right\vert ^{2}$, and the amplitudes $\Psi _{nm}$ in the bulk
satisfy the non-linear scattering equation (Sec. III in \cite{SM})%
\begin{equation}
S_{0}\Psi _{nm}=e^{-i\omega _{\mathrm{d}}L}e^{i\chi \mathcal{N}_{nm}L}\tilde{%
\Psi}_{nm}.  \label{NLE}
\end{equation}

The light distribution in the network is characterized by Eq. (\ref{NLE})
and the boundary conditions%
\begin{equation}
b_{u,0m}=\frac{a_{d,1m}}{ir_{\mathrm{BM}}}e^{-i\omega _{\mathrm{d}}L}-\frac{%
t_{\mathrm{BM}}}{ir_{\mathrm{BM}}}A_{\mathrm{in},m}^{(0)}e^{-i\omega _{%
\mathrm{d}}L/2}  \label{BC}
\end{equation}%
and $a_{d,N_{y}+1m}=a_{u,N_{y}m}$ on the top and bottom edges, respectively,
where $b_{u,0m}=(ia_{r,1m}+a_{l,1m})/\sqrt{2}$ and $t_{\mathrm{BM}}=\sqrt{%
1-r_{\mathrm{BM}}^{2}}$. The translational symmetry in Eq. (\ref{NLE}) and
boundary conditions give rise to $%
a_{(r,u,l,d),nm}=a_{(r,u,l,d),n}e^{ik_{x}m}/\sqrt{N_{x}}$.

\begin{figure}[tbp]
\includegraphics[bb=73 295 532 750, width=8 cm, clip]{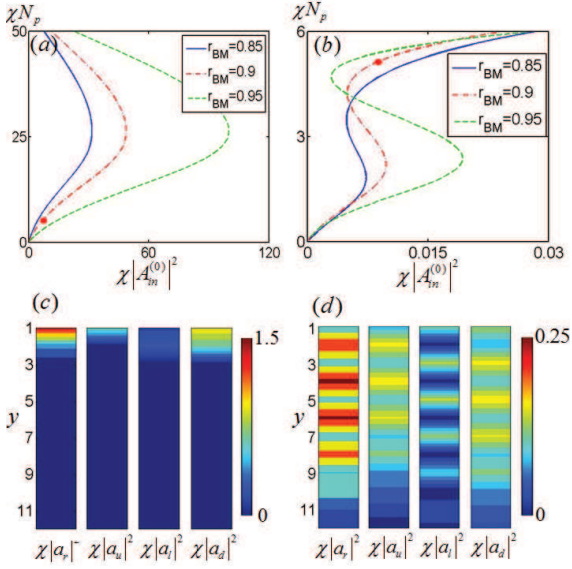}
\caption{(Color Online) Light distributions of the nonlinear system on the
cylinder, where the size is $24\times 12$, $\protect\theta _{0}=\protect\pi %
/2$, $k_{x}=0.26$, and $L$ is taken as unit. The relation of the total
intensity of (a) edge and (b) bulk modes in the network with different
reflection indices $r_{\mathrm{BM}}$ and the input intensity of driving
light with frequencies (a) $\protect\omega^{(1)}_{\mathrm{d}}=0.22$ and (b) $%
\protect\omega^{(2)}_{\mathrm{d}}=4.5\times 10^{-2}$. The stable internal
intensities of (c) edge and (d) bulk modes for $\left\vert \protect\chi%
\right\vert N_{\mathrm{p}}=5$ and $r_{\mathrm{BM}}=0.9$ (the red dots).}
\label{IDrelation}
\end{figure}

By numerically solving Eq. (\ref{NLE}), we show the total light intensity $%
N_{\mathrm{p}}=\sum_{n,s=r,u,l,d}\left\vert a_{s,n}\right\vert ^{2}$ versus
the driving strength $\left\vert A_{\mathrm{in}}^{(0)}\right\vert ^{2}$ in
the first row of Fig. \ref{IDrelation} for $\omega _{\mathrm{d}}^{(1)}\sim
0.22/L$ (Fig. \ref{IDrelation}a) and $\omega _{\mathrm{d}}^{(2)}\sim
4.5\times 10^{-2}/L$ (Fig. \ref{IDrelation}b), respectively, where $%
k_{x}\sim 0.26$ and the size of network is $N_{x}=24$, $N_{y}=12$. The $%
\left\vert \chi \right\vert N_{\mathrm{p}}$ versus $\left\vert \chi
\right\vert \left\vert A_{\mathrm{in}}^{(0)}\right\vert ^{2}$ curves display
that for the given parameters ($k_{x}$,$\omega _{\mathrm{d}}$), the driving
light with the amplitude $A_{\mathrm{in}}^{(0)}$ generates multiple light
intensities in the steady state of the network \cite{SM}. As discussed in
the following section (also Sec. IV in Ref. \cite{SM}), large domains of the
steady state solutions in Fig. \ref{IDrelation}a and b are unstable to small
perturbations. Here, we consider only stable solutions. The qualitative
origin of these complex stabilities can be traced to the behavior of a
single fiber segment with mirrors \cite{SM,instability3}.

For driving frequencies $\omega _{\mathrm{d}}^{(1)}$ and $\omega _{\mathrm{d}%
}^{(2)}$, Figs. \ref{IDrelation}c, d show that distinct light distribution $%
\left\vert \chi \right\vert \left\vert a_{(r,u,l,d),n}\right\vert ^{2}$ are
generated for interacting edge and bulk modes. Here, the total intensity $%
\left\vert \chi \right\vert N_{\mathrm{p}}=5/L$ and $r_{\mathrm{BM}}=0.9$.
We emphasize that the topologically protected chiral edge mode survives even
in the nonlinear regime, where the chirality is displayed by the dominating
right-moving intensity in Fig. \ref{IDrelation}c.

\textit{Bogoliubov excitations in non-linear optics.}--- Additional weak
probe light $\delta A_{m}^{\mathrm{in}}(t)=\sum_{\nu =\pm }\delta A_{\mathrm{%
in},m}^{(\nu )}e^{-i\nu \omega _{\mathrm{f}}t}$ through the top boundary
induces the fluctuations%
\begin{eqnarray}
\delta \phi _{s,nm}(x,t) &=&\frac{1}{\sqrt{N_{x}}}[\delta \psi
_{s,n,p_{x}}(x)e^{ip_{x}m}e^{-i\omega _{\mathrm{f}}t}  \notag \\
&&+\delta \psi _{s,n,q_{x}}(x)e^{iq_{x}m}e^{i\omega _{\mathrm{f}}t}]
\label{BF}
\end{eqnarray}%
around steady states, where the reflected fluctuation field\ from the
network is $\delta A_{m}^{\mathrm{out}}(t)=\sum_{\nu =\pm }\delta A_{\mathrm{%
out},m}^{(\nu )}e^{-i\nu \omega _{\mathrm{f}}t}$. Here, the positive- and
negative- frequency ($\pm \omega _{\mathrm{f}}$) components $\delta A_{%
\mathrm{in(out)},m}^{(\pm )}=e^{ik_{x}m}\delta A_{\mathrm{in(out)}}^{(\pm
)}e^{\pm i(p_{x}-k_{x})m}/\sqrt{N_{x}}$ have the quasi-momenta $p_{x}$ and $%
q_{x}=2k_{x}-p_{x}$, respectively.

By linearization of nonlinear motion equations around the steady state in
each fiber \cite{instability1,instability2,instability3}, the $S$-matrix of
nodes and the boundary conditions of the cylindrical network result in the
input-output relation $(\delta A_{\mathrm{out}}^{(+)},\delta A_{\mathrm{out}%
}^{(-)\ast })^{T}=M_{\mathrm{IO}}(\delta A_{\mathrm{in}}^{(+)},\delta A_{%
\mathrm{in}}^{(-)\ast })^{T}$ and the scattering equation (Sec. IV in \cite%
{SM})%
\begin{equation}
\mathbf{D}(\omega _{\mathrm{f}})\left(
\begin{array}{c}
\delta \mathbf{a}_{p_{x}} \\
\delta \mathbf{a}_{q_{x}}^{\ast }%
\end{array}%
\right) =t_{\mathrm{BM}}\left(
\begin{array}{c}
\delta \mathbf{A}_{\mathrm{in}}^{(+)}e^{i(\omega _{\mathrm{d}}-\omega _{%
\mathrm{f}})\frac{L}{2}} \\
\delta \mathbf{A}_{\mathrm{in}}^{(-)\ast }e^{-i(\omega _{\mathrm{d}}+\omega
_{\mathrm{f}})\frac{L}{2}}%
\end{array}%
\right)
\end{equation}%
for the fluctuation fields $\delta \mathbf{a}_{p_{x}}=(\delta
a_{r,n,p_{x}},\delta a_{u,n,p_{x}},\delta a_{l,n,p_{x}},\delta
a_{d,n,p_{x}})^{T}$ around the steady state configuration $a_{s,n}$. Here, $%
\delta \mathbf{A}_{\mathrm{in}}^{(\pm )}=\delta A_{\mathrm{in}}^{(\pm )}(%
\mathbf{0},\mathbf{0},\mathbf{0},\mathbf{1})^{T}$ is composed of the null
vector $\mathbf{0}$ and $\mathbf{1}=(1,0,...,0)$, which are of dimension $%
N_{y}$. The steady-state is stable if all roots $\mathcal{E}_{f}$ of det$%
\mathbf{D}(\omega _{\mathrm{f}})$ satisfy Im$\mathcal{E}_{f}<0$.

\begin{figure}[tbp]
\includegraphics[bb=15 429 578 743, width=8 cm, clip]{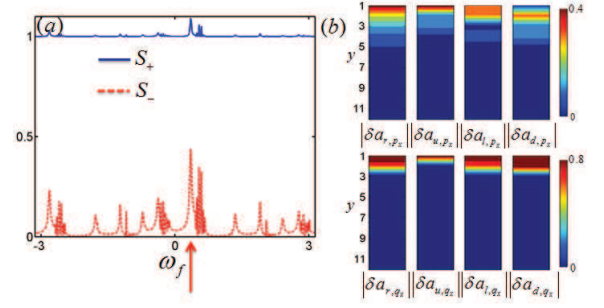}
\caption{(Color Online) Bogoliubov fluctuations on the cylinder, where the
system size is $24\times 12$, $\protect\theta _{0}=\protect\pi /2$. The
amplitude $|\protect\delta A_{\mathrm{in}}^{(+)}|$ of the probe light and $L$
are taken as unit. (a) The squeezing spectra of probe field with the
positive frequency, where the Bogoliubov fluctuations above the stable edge
steady states are generated; (b) The distributions of Bogoliubov edge
fluctuations above the stable edge steady states, where the Bogoliubov
excitation has the frequency $\protect\omega _{\mathrm{f}}=0.35$ as shown by
the red arrow in (a).}
\label{stability}
\end{figure}

The stability of the solutions in Fig. \ref{IDrelation} is analyzed by
solving det$\mathbf{D}(\mathcal{E}_{f})=0$, which shows that the steady
states in Fig. \ref{IDrelation}c, d are stable. Induced by the
\textquotedblleft condensation\textquotedblright\ $a_{s,n}$ in the steady
state, the fluctuation $\delta \mathbf{a}_{p_{x}}$ couples to the conjugate
amplitude $\delta \mathbf{a}_{q_{x}}^{\ast }$, known as the Bogoliubov
fluctuation. As a result, the probe light with positive frequency ($\delta
A_{\mathrm{in}}^{(-)}=0$) and momentum $p_{x}$ generates the Bogoliubov
fluctuations (\ref{BF}) with frequencies $\pm \omega _{\mathrm{f}}$, which
eventally gives rise to the emission of \textquotedblleft
classically\textquotedblright\ squeezed light $\delta A_{\mathrm{out},m}(t)$
\cite{QT}. The squeezing behavior is characterized by the squeezing spectra $%
S_{+}=\left\vert \delta A_{\mathrm{out}}^{(+)}/\delta A_{\mathrm{in}%
}^{(+)}\right\vert $ and $S_{-}=\left\vert \delta A_{\mathrm{out}%
}^{(-)}/\delta A_{\mathrm{in}}^{(+)}\right\vert $ of reflected light, where $%
S_{+}^{2}-S_{-}^{2}=1$ reflects the bosonic nature of light.

Around the stable edge steady state (Fig. \ref{IDrelation}c), the probe
field with momentum $p_{x}\sim 0.52$ induces the generation of squeezed
light with the spectra displayed in Fig. \ref{stability}a. The peak around
the frequency $\omega _{\mathrm{f}}\sim 0.35/L$ in the squeezing spectra is
the signature of the Bogoliubov fluctuation $\delta \mathbf{a}_{p_{x}}$ and $%
\delta \mathbf{a}_{q_{x}}^{\ast }$ in the network. As shown in Fig. \ref%
{stability}b, the large light distribution at the top boundary generates
strong coupling of Bogoliubov fluctuations localized at the edge, which
results in comparable magnitudes of $\delta \mathbf{a}_{p_{x}}$ and $\delta
\mathbf{a}_{q_{x}}^{\ast }$.

In the bulk steady state (Fig. \ref{IDrelation}d), edge fluctuations can
also be generated by the probe light with frequency $\omega _{\mathrm{f}%
}\sim 1.19/L$ and $p_{x}\sim 0.26$. However, in contrast to the Bogoliubov
excitation in the edge steady state in Fig. \ref{stability}, due to the
small light distribution along the boundary, the counterpart $\left\vert
\delta \mathbf{a}_{q_{x}}^{\ast }\right\vert \ll \left\vert \delta \mathbf{a}%
_{p_{x}}\right\vert $ of edge Bogoliubov fluctuations hardly response to the
driving field with positive frequency (e.g. $\left\vert \delta \mathbf{a}%
_{q_{x}}^{\ast }/\delta A_{\mathrm{in}}^{(+)}\right\vert \sim 10^{-3}$).

\textit{Conclusions}.--- We propose to realize QSH effects of classical
light with broad topological band gaps by optically passive elements in a 2D
network. For the linear network, we find robust helical edge modes in broad
topological band gaps, and investigate their generation and detection. In
the nonlinear network, the interplay of the Kerr nonlinearities and
topological effects gives rise to interacting edge (bulk) steady states and
squeezed edge fluctuations around these steady states.

The work was funded by the European Union Integrated project SIQS. HJK
acknowledges support as a Max Planck Institute for Quantum Optics
Distinguished Scholar, as well as funding from NSF Grant PHY-1205729, the
DOD NSSEFF program, the AFOSR QuMPASS MURI, the ONR QOMAND MURI, and the
IQIM, an NSF Physics Frontiers Center with support of the Moore Foundation.

\newpage \widetext

\begin{center}
\textbf{\large Supplemental Material}
\end{center}

\setcounter{equation}{0} \setcounter{figure}{0} 
\makeatletter

\renewcommand{\thefigure}{SM\arabic{figure}} \renewcommand{\thesection}{SM%
\arabic{section}} \renewcommand{\theequation}{SM\arabic{equation}}

This supplemental material is divided in four sections. In the first one,
Sec. SMI, we explain the construction of nodes, where two beamsplitters and
two perfect mirrors are applied to change the propagation directions of the
light in fibers. The analytical expression for the $S$-matrix for each node
is given explicitly.

The second section (Sec. SMII) is devoted to analyze the light propagation
in the fiber as a Kerr medium. After solving the nonlinear equations that
describe the light propagating in the horizontal and vertical fibers, in
Sec. SMIIA we investigate the properties of Bogoliubov fluctuations by
linearizing the nonlinear equations around the steady state solutions. To
get insight into the steady-state stability of the whole network, in Sec.
SMIIB we utilize a single nonlinear Fabry-Perot cavity as an paradigmatic
example to analyze the stability of steady states.

By combining the results of $S$-matrices at each node (Sec. SMI) and in
fibers (Sec. SMIIA), in Sec. SMIII we derive a nonlinear scattering equation
in the whole network with different geometries, which determines the steady
state properties. The dynamics of fluctuations in each fiber obey the\
linearized Bogoliubov equation (Sec. SMIIA), together with the $S$-matrix of
nodes, which results in a linear scattering equation of Bogoliubov
fluctuations in the network. The linear scattering equation is used to
analyze the stability of steady states in the network.

\section{SMI-Construction of nodes}

\begin{figure}[tbp]
\includegraphics[bb=24 162 580 760, width=10 cm, clip]{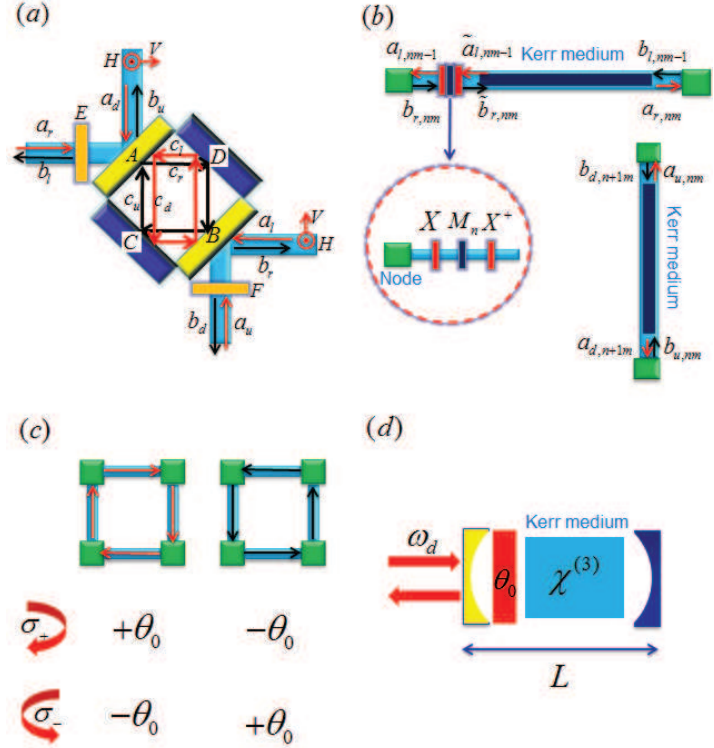}
\caption{(Color Online) Schematic for the nodes, the horizontal and vertical
fibers, the plaquette in the network, and a single Fabry-Perot cavity: (a) A
node connects horizontal and vertical links. Here, $a_{r,u,l,d} $ and $%
b_{r,u,l,d}$ denote the input and output amplitudes, and $c_{r,u,l,d}$
denote the amplitudes in the cavity. The horizontal and vertical
polarizations ($H$,$V$) in fibers are shown by the red dots and arrows. (b)
The horizontal and the vertical links with the Kerr medium and birefringent
elements, where the Kerr medium is put on the right hand side of three
birefringent elements in the horizontal link, where the birefringent
elements are assumed to have no Kerr nonlinearity ($\protect\chi=0$). From
left to right, the three birefringent elements are described by the Jones
matrices $X=e^{i\protect\pi \protect\sigma _{x}/4}$, $M_{n}=e^{in\protect%
\theta _{0}\protect\sigma _{z}}$, and $X^{\dagger }$ in the linear
polarization basis $(H,V)$. (c) The $\protect\sigma_{+}$ polarized light
acquires the phase $\protect\theta_0$ ($-\protect\theta_0$) by propagating
(anti-) clockwisely in each plaquette, while the $\protect\sigma_{-}$
polarized light acquires the phase $-\protect\theta_0$ ($\protect\theta_0$)
by propagating (anti-) clockwisely in each plaquette. (d) A single
Fabry-Perot cavity with Kerr nonlinearity and an anisotropic phase plate
placed next to the left end-mirror to mimic the horizontal link, where the
driving field with frequency $\protect\omega_d$ is applied.}
\label{nodefiber}
\end{figure}

In this section, we show the construction of the node by two beamsplitters
and two perfect mirrors, and derive the $S$-matrix of the node. Since the
polarizations of light are always orthogonal to the propagation direction,
the directions of vertical polarizations are different in the horizontal and
vertical fibers. As shown in Fig. \ref{nodefiber}a, the horizontal
polarizations in the fibers are chosen to be pointing out of the 2D plane,
while the vertical polarizations are pointing up and right in the horizontal
and vertical fibers, respectively.

As shown in Fig. \ref{nodefiber}a, the input and output amplitudes of the
node are $a_{r,u,l,d}$ and $b_{r,u,l,d}$, respectively. The input and output
amplitudes of the beamsplitter $A$ in the inner-cavity are $c_{u,l}$ and $%
c_{r,d}$, respectively. Elements $C$ and $D$ are perfectly reflecting
mirrors. For the beamsplitters $A$, $B$,\ the amplitudes obey the scattering
equations%
\begin{equation}
S_{\mathrm{bs}}^{(A)}\left(
\begin{array}{c}
\sigma _{z}a_{r} \\
c_{u} \\
c_{l} \\
a_{d}%
\end{array}%
\right) =\left(
\begin{array}{c}
c_{r} \\
b_{u} \\
\sigma _{z}b_{l} \\
c_{d}%
\end{array}%
\right) ;S_{\mathrm{bs}}^{(B)}\left(
\begin{array}{c}
c_{d} \\
\sigma _{z}a_{u} \\
a_{l} \\
c_{r}%
\end{array}%
\right) =\left(
\begin{array}{c}
b_{r} \\
c_{l} \\
c_{u} \\
\sigma _{z}b_{d}%
\end{array}%
\right) ,
\end{equation}%
where we assume that the size of the node is much smaller than the
wavelength of light such that the free propagation phase in the node can be
neglected, and the two-component amplitudes $a_{r,u,l,d}$, $b_{r,u,l,d}$,
and $c_{r,u,l,d}$ are defined in the polarization basis $(H,V)$. The Jones
matrix $\sigma _{z}$ is introduced to describe the two birefringent elements
($E,F$) in close proximity to the beamsplitters ($A,B$). The $S$-matrix $S_{%
\mathrm{bs}}^{(A)}=S_{\mathrm{bs}}^{(B)}=S_{\mathrm{bs}}$:%
\begin{equation}
S_{\mathrm{bs}}=\left(
\begin{array}{cccc}
t_{\mathrm{bs}} & ir_{\mathrm{bs}}\sigma _{z} & 0 & 0 \\
ir_{\mathrm{bs}}\sigma _{z} & t_{\mathrm{bs}} & 0 & 0 \\
0 & 0 & t_{\mathrm{bs}} & ir_{\mathrm{bs}}\sigma _{z} \\
0 & 0 & ir_{\mathrm{bs}}\sigma _{z} & t_{\mathrm{bs}}%
\end{array}%
\right)
\end{equation}%
for the beamsplitter is given by the real reflection and transmission
coefficients $r_{\mathrm{bs}}$ and $t_{\mathrm{bs}}=\sqrt{1-r_{\mathrm{bs}%
}^{2}}$, where the matrix $\sigma _{z}$ describes the direction change of
the vertical polarization due to the Fresnel reflection rule.

By elimination of the fields in the ring cavity formed by elements $A$, $B$,
$C$, $D$ and by incorporating the elements ($E,F$), we establish the
scattering equation for the node $S_{\mathrm{node}}\Psi _{a}=\Psi _{b}$,
where the $S$-matrix%
\begin{equation}
S_{\mathrm{node}}=\left(
\begin{array}{cccc}
0 & ir_{\mathrm{node}} & 0 & t_{\mathrm{node}} \\
ir_{\mathrm{node}} & 0 & t_{\mathrm{node}} & 0 \\
0 & t_{\mathrm{node}} & 0 & ir_{\mathrm{node}} \\
t_{\mathrm{node}} & 0 & ir_{\mathrm{node}} & 0%
\end{array}%
\right)
\end{equation}%
relates the input amplitudes $\Psi _{a}=(a_{r},a_{u},a_{l},a_{d})^{T}$ and
the output amplitudes $\Psi _{b}=(b_{r},b_{u},b_{l},b_{d})^{T}$. The
effective reflection and transmission coefficients of the node are $r_{%
\mathrm{node}}=2r_{\mathrm{bs}}/(1+r_{\mathrm{bs}}^{2})$ and $t_{\mathrm{node%
}}=t_{\mathrm{bs}}^{2}/(1+r_{\mathrm{bs}}^{2})$. In our scheme, $r_{\mathrm{%
node}}=t_{\mathrm{node}}=1/\sqrt{2}$ and the corresponding reflection
coefficient of the node is $r_{\mathrm{bs}}=\sqrt{2}-1$.

\section{SMII-Light propagation in fibers}

This section is divided in two subsections. In Sec. SMIIA, the propagation
of the light in a fiber with the non-linear Kerr medium is analyzed. In Sec.
SMIIB, a simple nonlinear system, i.e., a single Fabry-Perot cavity, is
analyzed, where the stability of steady states is investigated.

\subsection{SMIIA-Steady-state solutions and fluctuations in fibers}

As shown in Fig. \ref{nodefiber}b, without loss of generality, the three
birefringent elements are placed close to the node on the left side of the
horizontal fiber, and the Kerr medium is on the right hand side\ of these
birefringent elements. The oppositely directed effective magnetic fields for
circular polarized light $\sigma _{\pm }$ are generated by the
birefringences, as shown in Fig. \ref{nodefiber}c.

As shown by Eq. (\ref{S0}) in Sec. SMIII, two $\sigma _{\pm }$ polarized
light are decoupled in the linear network. Hence, in the analysis of
nonlinear system, we are able to treat the $\sigma _{\pm }$ polarizations
independently, as long as these polarizations are separately excited by the
external input (i.e., only $\sigma _{+}$ OR $\sigma _{-}$ polarization
circulating in the fiber links). Alternatively, by fiat we set to zero terms
related to cross phase modulation between orthogonal circular polarizations
in this first analysis of the nonlinear system, as discussed in Chapter 6 of
Ref. \cite{textbook} and Chapter 4 of Ref. \cite{NOP}.

In the Kerr medium of the fiber connecting nodes $(n,m)$ and $(n,m-1)$, the
right- and left- moving fields obey the motion equations \cite%
{instability1,instability2,instability3}%
\begin{eqnarray}
&&i\partial _{t}\phi _{r,nm}(x,t)+i\partial _{x}\phi _{r,nm}(x,t)  \notag \\
&=&\chi \lbrack \left\vert \phi _{r,nm}(x,t)\right\vert ^{2}+2\left\vert
\phi _{l,nm-1}(x,t)\right\vert ^{2}]\phi _{r,nm}(x,t),  \label{SMR}
\end{eqnarray}%
and%
\begin{eqnarray}
&&i\partial _{t}\phi _{l,nm-1}(x,t)-i\partial _{x}\phi _{l,nm-1}(x,t)  \notag
\\
&=&\chi \lbrack \left\vert \phi _{l,nm-1}(x,t)\right\vert ^{2}+2\left\vert
\phi _{r,nm}(x,t)\right\vert ^{2}]\phi _{l,nm-1}(x,t),  \label{SML}
\end{eqnarray}%
where $x$ is the distance along the fiber and the cross Kerr interaction of
fields with orthogonal polarizations is neglected.\

The formal solutions for Eqs. (\ref{SMR}) and (\ref{SML}) are%
\begin{eqnarray}
\phi _{r,nm}(x,t) &=&[a_{r,nm}+\delta \phi
_{r,nm}(x,t)]e^{ik_{r}(x-L)}e^{-i\omega t},  \notag \\
\phi _{l,nm-1}(x,t) &=&[\tilde{a}_{l,nm-1}+\delta \phi
_{l,nm-1}(x,t)]e^{-ik_{l}x}e^{-i\omega t},  \label{SL}
\end{eqnarray}%
where $\delta \phi _{r,nm}$ and $\delta \phi _{l,nm-1}$ are the fluctuations
around the steady state solutions $a_{r,nm}$ and $\tilde{a}_{l,nm-1}$. As
shown in Fig. \ref{nodefiber}b, $a_{r,nm}$ is the amplitude of the
right-moving input field to the node $(n,m)$, and $\tilde{a}_{l,nm-1}$ is
the amplitude of the left-moving input field to birefringent elements. The
wave vectors%
\begin{eqnarray}
k_{r} &=&\omega -\chi \lbrack \left\vert a_{r,nm}\right\vert
^{2}+2\left\vert \tilde{a}_{l,nm-1}\right\vert ^{2}],  \notag \\
k_{l} &=&\omega -\chi \lbrack \left\vert \tilde{a}_{l,nm-1}\right\vert
^{2}+2\left\vert a_{r,nm}\right\vert ^{2}],
\end{eqnarray}%
of the right- and left-moving fields are given by the intensities of the
fields and the characteristic frequency $\omega $, where $\omega =\mathcal{E}
$ is the eigenfrequency of the closed network, and $\omega =\omega _{\mathrm{%
d}}$ is the frequency of the driving field applied to the open network.

The steady state solution establishes the relations%
\begin{eqnarray}
e^{-in\sigma \theta _{0}}b_{r,nm} &=&\tilde{b}_{r,nm}=e^{-ik_{r}L}a_{r,nm},
\notag \\
e^{in\sigma \theta _{0}}b_{l,nm-1} &=&e^{-ik_{l}L}e^{in\sigma \theta _{0}}%
\tilde{a}_{l,nm-1}=e^{-ik_{l}L}a_{l,nm-1},  \label{Hss}
\end{eqnarray}%
where, as shown in Fig. \ref{nodefiber}b, $b_{r,nm}$ is the amplitude of the
right-moving output field from the node $(n,m-1)$, $\tilde{b}_{r,nm}$ is the
amplitude of the right-moving on the right hand side of birefringent
elements, $b_{l,nm-1}$ is the amplitude of the left-moving output field from
the node $(n,m)$, $a_{l,nm-1}$ is the amplitude of the left-moving input
field to the node $(n,m-1)$, and $\sigma =\pm 1$ denotes the two orthogonal
circular polarizations.

Around the steady state solution, the motion equations (\ref{SMR}) and (\ref%
{SML}) are linearized as%
\begin{equation}
i\partial _{t}\delta \Psi +\mathbf{\Sigma }\partial _{x}\delta \Psi =\mathbf{%
M}_{H}\delta \Psi
\end{equation}%
for the Bogoliubov fluctuation $\delta \Psi =(\delta \phi _{r,nm},\delta
\phi _{l,nm-1},\delta \phi _{r,nm}^{\ast },\delta \phi _{l,nm-1}^{\ast
})^{T} $, where the matrices are%
\begin{equation}
\mathbf{\Sigma }=i\left(
\begin{array}{cccc}
1 & 0 & 0 & 0 \\
0 & -1 & 0 & 0 \\
0 & 0 & 1 & 0 \\
0 & 0 & 0 & -1%
\end{array}%
\right) ,
\end{equation}%
and%
\begin{equation}
\mathbf{M}_{H}\mathbf{=}\chi \left(
\begin{array}{cccc}
\left\vert a_{r,nm}\right\vert ^{2} & 2\tilde{a}_{l,nm-1}^{\ast }a_{r,nm} &
a_{r,nm}^{2} & 2a_{r,nm}\tilde{a}_{l,nm-1} \\
2a_{r,nm}^{\ast }\tilde{a}_{l,nm-1} & \left\vert \tilde{a}%
_{l,nm-1}\right\vert ^{2} & 2a_{r,nm}\tilde{a}_{l,nm-1} & \tilde{a}%
_{l,nm-1}^{2} \\
-a_{r,nm}^{\ast 2} & -2a_{r,nm}^{\ast }\tilde{a}_{l,nm-1}^{\ast } &
-\left\vert a_{r,nm}\right\vert ^{2} & -2a_{r,nm}^{\ast }\tilde{a}_{l,nm-1}
\\
-2a_{r,nm}^{\ast }\tilde{a}_{l,nm-1}^{\ast } & -\tilde{a}_{l,nm-1}^{\ast 2}
& -2\tilde{a}_{l,nm-1}^{\ast }a_{r,nm} & -\left\vert \tilde{a}%
_{l,nm-1}\right\vert ^{2}%
\end{array}%
\right) .  \label{MH}
\end{equation}%
The Bogoliubov mode $\delta \Psi =\delta \psi e^{-i\omega _{\mathrm{f}}t}$
with the fluctuation frequency $\omega _{\mathrm{f}}$ around $\omega $ obeys
the equation%
\begin{equation}
\omega _{\mathrm{f}}\delta \psi +\mathbf{\Sigma }\partial _{x}\delta \psi =%
\mathbf{M}_{H}\delta \psi ,  \label{flh}
\end{equation}%
where the time-independent field $\delta \psi =(\delta \psi _{r,nm},\delta
\psi _{l,nm-1},\delta \psi _{r,nm}^{\ast },\delta \psi _{l,nm-1}^{\ast
})^{T} $.

The formal solution of Eq. (\ref{flh}) leads to the relation%
\begin{equation}
e^{\mathbf{\Sigma }(\omega _{\mathrm{f}}-\mathbf{M}_{H})L}\left(
\begin{array}{c}
e^{-in\sigma \theta _{0}}\delta b_{r,nm} \\
e^{-in\sigma \theta _{0}}\delta a_{l,nm-1} \\
e^{in\sigma \theta _{0}}\delta b_{r,nm}^{\ast } \\
e^{in\sigma \theta _{0}}\delta a_{l,nm-1}^{\ast }%
\end{array}%
\right) =\left(
\begin{array}{c}
\delta a_{r,nm} \\
\delta b_{l,nm-1} \\
\delta a_{r,nm}^{\ast } \\
\delta b_{l,nm-1}^{\ast }%
\end{array}%
\right)  \label{Hf}
\end{equation}%
of the input and output fluctuation fields%
\begin{equation}
\delta a_{r,nm}=\delta \psi _{r,nm}(L),\delta a_{l,nm-1}=e^{in\sigma \theta
_{0}}\delta \psi _{l,nm-1}(0),
\end{equation}%
and%
\begin{equation}
\delta b_{r,nm}=e^{in\sigma \theta _{0}}\delta \psi _{r,nm}(0),\delta
b_{l,nm-1}=\delta \psi _{l,nm-1}(L),
\end{equation}%
around the steady state amplitudes $a_{r,nm}$ ($\delta a_{l,nm-1}$) and $%
b_{r,nm}$ ($b_{l,nm-1}$).

The same analysis is applied to light propagation in the vertical fiber
connecting nodes $(n,m)$ and $(n+1,m)$. The relations for the amplitudes of
the input fields $a_{u,nm},a_{d,n+1m}$ and the output field $%
b_{u,nm},b_{d,n+1m}$ (as shown in Fig. \ref{nodefiber}b) in the steady state
are%
\begin{equation}
b_{u,nm}=e^{-ik_{u}L}a_{u,nm},b_{d,n+1m}=e^{-ik_{d}L}a_{d,n+1m},  \label{Vss}
\end{equation}%
where the wavevectors are%
\begin{eqnarray}
k_{u} &=&\omega -\chi \lbrack \left\vert a_{u,nm}\right\vert
^{2}+2\left\vert a_{d,n+1m}\right\vert ^{2}],  \notag \\
k_{d} &=&\omega -\chi \lbrack \left\vert a_{d,n+1m}\right\vert
^{2}+2\left\vert a_{u,nm}\right\vert ^{2}].
\end{eqnarray}

The relation%
\begin{equation}
e^{\mathbf{\Sigma }(\omega _{\mathrm{f}}-\mathbf{M}_{V})L}\left(
\begin{array}{c}
\delta b_{u,nm} \\
\delta a_{d,n+1m} \\
\delta b_{u,nm}^{\ast } \\
\delta a_{d,n+1m}^{\ast }%
\end{array}%
\right) =\left(
\begin{array}{c}
\delta a_{u,nm} \\
\delta b_{d,n+1m} \\
\delta a_{u,nm}^{\ast } \\
\delta b_{d,n+1m}^{\ast }%
\end{array}%
\right)  \label{Vf}
\end{equation}%
relates the input and output fluctuation fields $\delta a_{u,nm}$ ($\delta
a_{d,n+1m}$) and $\delta b_{u,nm}$ ($\delta b_{d,n+1m}$) around the steady
state amplitudes $a_{u,nm}$ ($a_{d,n+1m}$) and $b_{u,nm}$ ($b_{d,n+1m}$),
where%
\begin{equation}
\mathbf{M}_{V}\mathbf{=}\chi \left(
\begin{array}{cccc}
\left\vert a_{u,nm}\right\vert ^{2} & 2a_{d,n+1m}^{\ast }a_{u,nm} &
a_{u,nm}^{2} & 2a_{u,nm}a_{d,n+1m} \\
2a_{u,nm}^{\ast }a_{d,n+1m} & \left\vert a_{d,n+1m}\right\vert ^{2} &
2a_{u,nm}a_{d,n+1m} & a_{d,n+1m}^{2} \\
-a_{u,nm}^{\ast 2} & -2a_{u,nm}^{\ast }a_{d,n+1m}^{\ast } & -\left\vert
a_{u,nm}\right\vert ^{2} & -2a_{u,nm}^{\ast }a_{d,n+1m} \\
-2a_{u,nm}^{\ast }a_{d,n+1m}^{\ast } & -a_{d,n+1m}^{\ast 2} &
-2a_{d,n+1m}^{\ast }a_{u,nm} & -\left\vert a_{d,n+1m}\right\vert ^{2}%
\end{array}%
\right) .  \label{MV}
\end{equation}

\subsection{SMIIB-Nonlinear Fabry-Perot cavity}

Before studying the steady state properties and the stability of the light
in the whole network, we use a paradigmatic example, i.e., the single
Fabry-Perot cavity with nonlinear Kerr medium \cite{instability3} to show
the stability analysis of steady states. Our goal is to understand better
the stability analysis for more complex 2D arrays of nonlinear fibers and
beam splitters.

As shown in Fig. \ref{nodefiber}d, the cavity with a perfect right
end-mirror is driven by the light with frequency $\omega _{\mathrm{d}}$
through a partially transmissive mirror at the left end. In the cavity, the
phase plate is placed next to the transmissive mirror. In propagation from
left to right, the light acquires the phase factor $e^{-i\theta _{0}}$.
Here, $\theta _{0}\neq 0$ ($\theta _{0}=0$) corresponds to the single
horizontal (vertical) fiber in the network.

The relations%
\begin{equation}
e^{-i\theta _{0}}b_{r}=e^{-ik_{r}L}a_{r},e^{i\theta
_{0}}b_{l}=e^{-ik_{l}L}a_{l}  \label{singlea}
\end{equation}%
of input $a_{r}$ ($a_{l}$) and output amplitude $b_{r}$ ($b_{l}$) follow
from Eq. (\ref{Hss}), where $L$ is the cavity length, and%
\begin{eqnarray}
k_{r} &=&\omega _{\mathrm{d}}-\chi (\left\vert a_{r}\right\vert
^{2}+2\left\vert a_{l}\right\vert ^{2}),  \notag \\
k_{l} &=&\omega _{\mathrm{d}}-\chi (\left\vert a_{l}\right\vert
^{2}+2\left\vert a_{r}\right\vert ^{2}).
\end{eqnarray}%
At the end-mirrors, the boundary conditions are $a_{r}=b_{l}$, and%
\begin{equation}
\left(
\begin{array}{c}
b_{r} \\
A_{\mathrm{out}}^{(0)}%
\end{array}%
\right) =\left(
\begin{array}{cc}
t_{\mathrm{BM}} & ir_{\mathrm{BM}} \\
ir_{\mathrm{BM}} & t_{\mathrm{BM}}%
\end{array}%
\right) \left(
\begin{array}{c}
A_{\mathrm{in}}^{(0)} \\
a_{l}%
\end{array}%
\right) ,  \label{singleb}
\end{equation}%
where $t_{\mathrm{BM}}$ ($r_{\mathrm{BM}}$) is the real transmission
(reflection) coefficients of the left end-mirror, and $A_{\mathrm{in}}^{(0)}$
($A_{\mathrm{out}}^{(0)}$) is the input (output) amplitude of the cavity.

By eliminating the output amplitude $b_{r}$ ($b_{l}$) in Eqs. (\ref{singlea}%
) and (\ref{singleb}), we obtain the nonlinear equation%
\begin{equation}
a_{r}=\frac{t_{\mathrm{BM}}A_{\mathrm{in}}^{(0)}e^{-i\theta _{0}}}{%
e^{-ikL}-ir_{\mathrm{BM}}e^{ikL}},  \label{singlenl}
\end{equation}%
that determines the amplitude $a_{r}=\left\vert a_{r}\right\vert e^{i\theta
_{r}}$, where $k_{r}=k_{l}\equiv k=\omega _{\mathrm{d}}-3\chi \left\vert
a_{r}\right\vert ^{2}$, and the output amplitude%
\begin{equation}
A_{\mathrm{out}}^{(0)}=ir_{\mathrm{BM}}A_{\mathrm{in}}^{(0)}+t_{\mathrm{BM}%
}a_{l}=\frac{e^{ikL}+ir_{\mathrm{BM}}e^{-ikL}}{e^{-ikL}-ir_{\mathrm{BM}%
}e^{ikL}}A_{\mathrm{in}}^{(0)}
\end{equation}%
of the cavity is determined by the relation $a_{l}=e^{i\theta
_{0}}e^{ikL}a_{r}$. In the good cavity limit $t_{\mathrm{BM}}\rightarrow 0$,
Eq. (\ref{singlenl}) determines the intensity-dependent frequency%
\begin{equation}
\mathcal{E}_{n}=\frac{n\pi }{L}-\frac{\pi }{4L}+3\chi \left\vert
a_{r}\right\vert ^{2}  \label{eclose}
\end{equation}%
of the closed cavity, where $n$ is an integer.

\begin{figure}[tbp]
\includegraphics[bb=22 246 569 745, width=10 cm, clip]{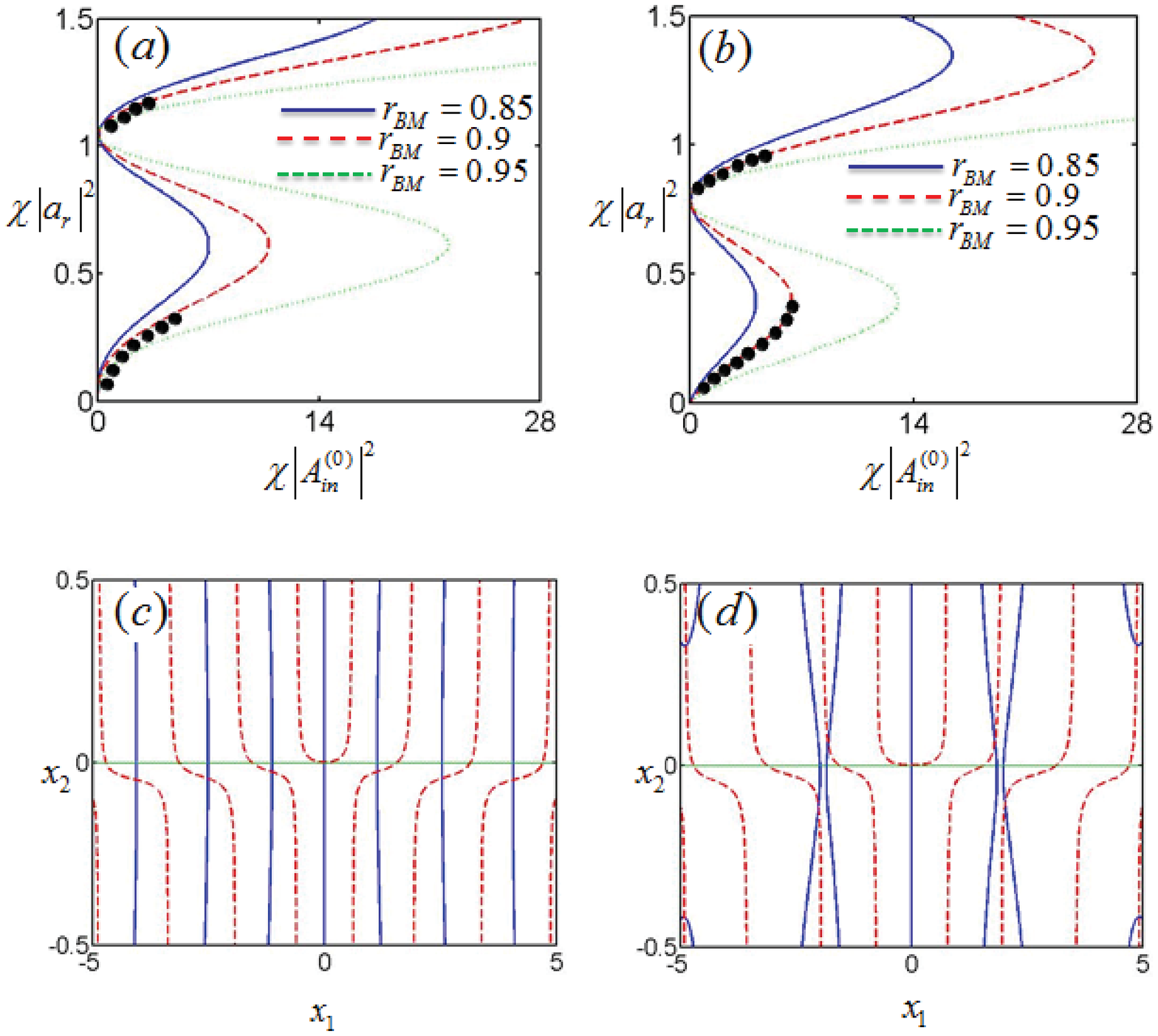}
\caption{(Color Online) Steady state solutions and stability analysis, where
$L$ is taken as unit. (a)-(b): The relation of the light intensity $\protect%
\chi \left\vert a_{r}\right\vert ^{2}$ in the cavity and the driving field
intensity $\protect\chi \left\vert A_{\mathrm{in}}^{(0)}\right\vert ^{2}$ in
steady state for the driving frequencies $\protect\omega_d=3\protect\pi/4$
(a) and $\protect\omega_d=\protect\pi/2$ (b). Here, the stable regimes are
marked by the black dots. (c)-(d): For $\protect\omega_d=3\protect\pi/4$,
the first and second equations in (\protect\ref{TC}) are shown by the solid
(blue) and dashed (red) curves, where $\protect\chi\left\vert A_{\mathrm{in}%
}\right\vert ^{2}=1$, $\protect\chi \left\vert a_{r}\right\vert ^{2}=1.12$
(c) and $\protect\chi\left\vert A_{\mathrm{in}}\right\vert ^{2}=5$, $\protect%
\chi \left\vert a_{r}\right\vert ^{2}=1.21$ (d).}
\label{FP}
\end{figure}

For different driving frequency $\omega _{\mathrm{d}}$, the relation%
\begin{equation}
x=\frac{y}{1-r_{\mathrm{BM}}^{2}}[1+r_{\mathrm{BM}}^{2}+2r_{\mathrm{BM}}\sin
(2\omega _{\mathrm{d}}L-6y)]
\end{equation}%
of $y=\chi \left\vert a_{r}\right\vert ^{2}$ and the input intensity $x=\chi
\left\vert A_{\mathrm{in}}^{(0)}\right\vert ^{2}$ is shown in Fig. \ref{FP}a
and \ref{FP}b, where $L=1$ is taken as unit and $r_{\mathrm{BM}%
}=0.85,0.9,0.95$. When the driving frequency $\omega _{\mathrm{d}}$ is
resonant with the intrinsic frequency $\mathcal{E}_{n}$ of the closed
cavity, the output field $A_{\mathrm{out}}^{(0)}=-iA_{\mathrm{in}}^{(0)}$.

The figures \ref{FP}a and \ref{FP}b show that for a given $\omega _{\mathrm{d%
}}$, the driving field with a fixed intensity $\left\vert A_{\mathrm{in}%
}^{(0)}\right\vert ^{2}$ can generate multiple intracavity intensities. In
order to analyze the stability of these multiple steady states, we
investigate the energy spectrum of Bogoliubov fluctuations. It follows from
Eq. (\ref{Hf}) that the fluctuation fields satisfy%
\begin{equation}
e^{\mathbf{\Sigma }(\omega _{\mathrm{f}}-U_{\mathrm{s}}^{\dagger }\mathbf{M}%
_{\mathrm{s}}U_{\mathrm{s}})L}\left(
\begin{array}{c}
e^{-i\theta _{0}}\delta b_{r} \\
e^{-i\theta _{0}}\delta a_{l} \\
e^{i\theta _{0}}\delta b_{r}^{\ast } \\
e^{i\theta _{0}}\delta a_{l}^{\ast }%
\end{array}%
\right) =\left(
\begin{array}{c}
\delta a_{r} \\
\delta b_{l} \\
\delta a_{r}^{\ast } \\
\delta b_{l}^{\ast }%
\end{array}%
\right) ,  \label{Hs}
\end{equation}%
where the matrix%
\begin{equation}
\mathbf{M}_{\mathrm{s}}=\chi \left\vert a_{r}\right\vert ^{2}\left(
\begin{array}{cccc}
1 & 2e^{-ikL} & 1 & 2e^{ikL} \\
2e^{ikL} & 1 & 2e^{ikL} & e^{2ikL} \\
-1 & -2e^{-ikL} & -1 & -2e^{ikL} \\
-2e^{-ikL} & -e^{-2ikL} & -2e^{-ikL} & -1%
\end{array}%
\right)
\end{equation}%
for the single fiber, and the unitary matrix $U_{\mathrm{s}}=I_{2}\oplus
e^{2i\theta _{r}}I_{2}$ is determined by the two-dimensional identity matrix
$I_{2}$.

The fluctuation equation (\ref{Hs}) leads to the relation $\delta \mathbf{B}=%
\tilde{U}_{\theta }U_{b}^{-1}U_{a}U_{\theta }\delta \mathbf{A}$ of $\delta
\mathbf{B}=(\delta b_{r},\delta b_{l},\delta b_{r}^{\ast },\delta
b_{l}^{\ast })^{T}$ and $\delta \mathbf{A}=(\delta a_{r},\delta a_{l},\delta
a_{r}^{\ast },\delta a_{l}^{\ast })^{T}$, where the matrices%
\begin{equation}
U_{b}=\left(
\begin{array}{cccc}
P_{\mathrm{s},11} & 0 & P_{\mathrm{s},13} & 0 \\
-P_{\mathrm{s},21} & 1 & -P_{\mathrm{s},23} & 0 \\
P_{\mathrm{s},31} & 0 & P_{\mathrm{s},33} & 0 \\
-P_{\mathrm{s},41} & 0 & -P_{\mathrm{s},43} & 1%
\end{array}%
\right) ,U_{a}=\left(
\begin{array}{cccc}
1 & -P_{\mathrm{s},12} & 0 & -P_{\mathrm{s},14} \\
0 & P_{\mathrm{s},22} & 0 & P_{\mathrm{s},24} \\
0 & -P_{\mathrm{s},32} & 1 & -P_{\mathrm{s},34} \\
0 & P_{\mathrm{s},42} & 0 & P_{\mathrm{s},44}%
\end{array}%
\right)
\end{equation}%
are determined by the propagating matrix $P_{\mathrm{s}}=e^{\mathbf{\Sigma }%
(\omega _{\mathrm{f}}-U_{\mathrm{s}}^{\dagger }\mathbf{M}_{\mathrm{s}}U_{%
\mathrm{s}})L}$, and the diagonal matrices%
\begin{equation}
U_{\theta }=\left(
\begin{array}{cccc}
1 & 0 & 0 & 0 \\
0 & e^{-i\theta _{0}} & 0 & 0 \\
0 & 0 & 1 & 0 \\
0 & 0 & 0 & e^{i\theta _{0}}%
\end{array}%
\right) ,\tilde{U}_{\theta }=\left(
\begin{array}{cccc}
e^{i\theta _{0}} & 0 & 0 & 0 \\
0 & 1 & 0 & 0 \\
0 & 0 & e^{-i\theta _{0}} & 0 \\
0 & 0 & 0 & 1%
\end{array}%
\right) .
\end{equation}

On the other hand, the boundary conditions at the end mirrors are $\delta
a_{r}=\delta b_{l}e^{-ikL}$ and%
\begin{equation}
\left(
\begin{array}{c}
\delta b_{r}e^{-ikL} \\
\delta A_{\mathrm{out}}%
\end{array}%
\right) =\left(
\begin{array}{cc}
t_{\mathrm{BM}} & ir_{\mathrm{BM}} \\
ir_{\mathrm{BM}} & t_{\mathrm{BM}}%
\end{array}%
\right) \left(
\begin{array}{c}
\delta A_{\mathrm{in}} \\
\delta a_{l}%
\end{array}%
\right) .
\end{equation}%
By eliminating the fluctuation field $\delta B$, we obtain the scattering
equation%
\begin{equation}
(\tilde{U}_{\theta }U_{b}^{-1}U_{a}U_{\theta }-U_{k}R_{\mathrm{BM}})\delta
\mathbf{A}=t_{\mathrm{BM}}U_{k}\delta \mathbf{A}_{d}
\end{equation}%
with the driving term $\delta \mathbf{A}_{d}=(\delta A_{\mathrm{in}%
},0,\delta A_{\mathrm{in}}^{\ast },0)^{T}$, where the matrices $%
U_{k}=e^{ikL}I_{2}\oplus e^{-ikL}I_{2}$ and%
\begin{equation}
R_{\mathrm{BM}}=\left(
\begin{array}{cccc}
0 & ir_{\mathrm{BM}} & 0 & 0 \\
1 & 0 & 0 & 0 \\
0 & 0 & 0 & -ir_{\mathrm{BM}} \\
0 & 0 & 1 & 0%
\end{array}%
\right) .
\end{equation}%
The zeros $D(\mathcal{E}_{\mathrm{f}})=0$ of the determinant%
\begin{equation}
D(\omega _{\mathrm{f}})=\det (\tilde{U}_{\theta }U_{b}^{-1}U_{a}U_{\theta
}-U_{k}R_{\mathrm{BM}})
\end{equation}%
determines the stability of the steady state solution, where the steady
state is stable if all Im$\mathcal{E}_{\mathrm{f}}<0$.

For the good cavity limit $t_{\mathrm{BM}}\rightarrow 0$, the momentum $%
kL=n\pi -\pi /4$, and the eigenfrequency of Bogoliubov fluctuations is $%
\mathcal{E}_{\mathrm{f}}=n_{\mathrm{f}}\pi $, where $n_{\mathrm{f}}$ is an
integer. For the open cavity, the condition $D(\mathcal{E}_{\mathrm{f}})=0$
leads to the two transcendental equations%
\begin{eqnarray}
\text{Re}D(x_{1}+ix_{2}) &=&0,  \notag \\
\text{Im}D(x_{1}+ix_{2}) &=&0,  \label{TC}
\end{eqnarray}%
for $x_{1}=$Re$\mathcal{E}_{\mathrm{f}}$ and $x_{2}=$Im$\mathcal{E}_{\mathrm{%
f}}$. In Figs. \ref{FP}c and \ref{FP}d, we show the two curves given by Eq. (%
\ref{TC}) for different driving intensities $\chi \left\vert A_{\mathrm{in}%
}^{(0)}\right\vert ^{2}$, where the intersections of two curves determines
the solution $x_{1}$ and $x_{2}$. As shown in Fig. \ref{FP}d, the positive
coordinates $x_{2}>0$ at points of intersection imply an unstable steady
state. In Figs. \ref{FP}a and \ref{FP}b, the stable regimes are marked by
the black dots, where these stable solutions are in the positive slope
regimes of $\chi \left\vert a_{r}\right\vert ^{2}$ versus $\chi \left\vert
A_{\mathrm{in}}^{(0)}\right\vert ^{2}$ curves.

\section{SMIII-Scattering equations on different geometries}

In this section, we use the results (\ref{Hss}) and (\ref{Vss}) to derive
the scattering equation for the amplitude $a_{r,u,l,d}$ in the steady state
of the whole network. Here, in terms of different boundary conditions, we
analyze the scattering equations describing the closed and open networks on
the torus, cylindrical, and planar geometries.

The combination of Eqs. (\ref{Hss}), (\ref{Vss}) and the $S$-matrix $S_{%
\mathrm{node}}$ leads to the scattering equation%
\begin{equation}
S_{0}\left(
\begin{array}{c}
a_{r,nm} \\
a_{u,nm} \\
a_{l,nm} \\
a_{d,nm}%
\end{array}%
\right) =e^{-i\omega L}e^{i\chi \mathcal{N}_{nm}L}\left(
\begin{array}{c}
a_{r,nm+1} \\
a_{u,n-1m} \\
a_{l,nm-1} \\
a_{d,n+1m}%
\end{array}%
\right)  \label{SN}
\end{equation}%
for the input amplitudes of nodes in the bulk, where the $S$-matrix%
\begin{equation}
S_{0}=\frac{1}{\sqrt{2}}\left(
\begin{array}{cccc}
0 & ie^{-in\sigma \theta _{0}} & 0 & e^{-in\sigma \theta _{0}} \\
i & 0 & 1 & 0 \\
0 & e^{in\sigma \theta _{0}} & 0 & ie^{in\sigma \theta _{0}} \\
1 & 0 & i & 0%
\end{array}%
\right)  \label{S0}
\end{equation}%
describes the free propagation in the linear network, and the additional
phase induced by the Kerr nonlinearity is depicted by the intensity matrix%
\begin{equation}
\mathcal{N}_{nm}=\left(
\begin{array}{cccc}
\left\vert a_{r,nm+1}\right\vert ^{2}+2\left\vert a_{l,nm}\right\vert ^{2} &
0 & 0 & 0 \\
0 & \left\vert a_{u,n-1m}\right\vert ^{2}+2\left\vert a_{d,nm}\right\vert
^{2} & 0 & 0 \\
0 & 0 & \left\vert a_{l,nm-1}\right\vert ^{2}+2\left\vert
a_{r,nm}\right\vert ^{2} & 0 \\
0 & 0 & 0 & \left\vert a_{d,n+1m}\right\vert ^{2}+2\left\vert
a_{u,nm}\right\vert ^{2}%
\end{array}%
\right) .
\end{equation}%
Without the Kerr nonlinearity, i.e., $\chi =0$, the scattering matrix (\ref%
{S0}) shows that the $\sigma _{\pm }$ polarizations are decoupled.

\subsection{SMIIIA-Closed network}

For the network on the torus, the boundary condition is%
\begin{eqnarray}
a_{r,nN_{x}+1} &=&a_{r,n1},a_{l,n0}=a_{l,nN_{x}},  \notag \\
a_{u,0m} &=&a_{u,N_{y}m},a_{d,N_{y}+1m}=a_{d,1m},  \label{torus}
\end{eqnarray}%
where the size of the network is $N_{x}\times N_{y}$. For the network on
cylinder with the periodic boundary condition along $x$-direction,%
\begin{eqnarray}
a_{r,nN_{x}+1} &=&a_{r,n1},a_{l,n0}=a_{l,nN_{x}},  \notag \\
a_{u,0m} &=&a_{d,1m},a_{d,N_{y}+1m}=a_{u,N_{y}m}.  \label{cylinder}
\end{eqnarray}

For the planar network with boundary mirrors of unit reflectivity, the
boundary condition is%
\begin{eqnarray}
a_{r,nN_{x}+1} &=&a_{l,nN_{x}},a_{l,n0}=a_{r,n1},  \notag \\
a_{u,0m} &=&a_{d,1m},a_{d,N_{y}+1m}=a_{u,N_{y}m}.  \label{plane}
\end{eqnarray}%
Due to the translational symmetry of the network on torus and cylinder, the
solution has the form%
\begin{equation}
\left(
\begin{array}{c}
a_{r,nm} \\
a_{u,nm} \\
a_{l,nm} \\
a_{d,nm}%
\end{array}%
\right) =\frac{1}{\sqrt{N_{x}}}e^{ik_{x}m}\left(
\begin{array}{c}
a_{r,n} \\
a_{u,n} \\
a_{l,n} \\
a_{d,n}%
\end{array}%
\right) ,  \label{NS}
\end{equation}%
with the quasi-momentum $k_{x}=-\pi +2\pi n/N_{x}$ ($n=0,1,2,...,N_{x}-1$),
and the scattering Eq. (\ref{SN}) becomes%
\begin{equation}
S_{0}(k_{x})\left(
\begin{array}{c}
a_{r,n} \\
a_{u,n} \\
a_{l,n} \\
a_{d,n}%
\end{array}%
\right) =e^{-i\mathcal{E}L}e^{i\frac{\chi }{N_{x}}\mathcal{N}_{n}L}\left(
\begin{array}{c}
a_{r,n} \\
a_{u,n-1} \\
a_{l,n} \\
a_{d,n+1}%
\end{array}%
\right) .
\end{equation}%
where%
\begin{equation}
S_{0}(k_{x})=\frac{1}{\sqrt{2}}\left(
\begin{array}{cccc}
0 & ie^{-ik_{x}}e^{-in\sigma \theta _{0}} & 0 & e^{-ik_{x}}e^{-in\sigma
\theta _{0}} \\
i & 0 & 1 & 0 \\
0 & e^{ik_{x}}e^{in\sigma \theta _{0}} & 0 & ie^{ik_{x}}e^{in\sigma \theta
_{0}} \\
1 & 0 & i & 0%
\end{array}%
\right) ,
\end{equation}%
and the intensity matrix%
\begin{equation}
\mathcal{N}_{n}=\left(
\begin{array}{cccc}
\left\vert a_{r,n}\right\vert ^{2}+2\left\vert a_{l,n}\right\vert ^{2} & 0 &
0 & 0 \\
0 & \left\vert a_{u,n-1}\right\vert ^{2}+2\left\vert a_{d,n}\right\vert ^{2}
& 0 & 0 \\
0 & 0 & \left\vert a_{l,n}\right\vert ^{2}+2\left\vert a_{r,n}\right\vert
^{2} & 0 \\
0 & 0 & 0 & \left\vert a_{d,n+1}\right\vert ^{2}+2\left\vert
a_{u,n}\right\vert ^{2}%
\end{array}%
\right) .
\end{equation}

By taking into account the boundary conditions (\ref{torus}) and (\ref%
{cylinder}), the scattering equation for the entire network on the torus and
the cylinder can be symbolically written as%
\begin{equation}
S_{0}(k_{x})\mathbf{a}=e^{-i\mathcal{E}L}e^{i\frac{\chi }{N_{x}}\mathcal{N}%
_{n}L}\mathbf{a}  \label{cclose}
\end{equation}%
in the basis $\mathbf{a}%
=(a_{r,1},...,a_{r,N_{y}};a_{u,1},...,a_{u,N_{y}};a_{l,1},...,a_{l,N_{y}};a_{d,1},...,a_{d,N_{y}})^{T}
$. Similarly, by the boundary condition (\ref{plane}), the scattering
equation for the planar network reads%
\begin{equation}
S_{0}\mathbf{a}=e^{-i\mathcal{E}L}e^{i\chi \mathcal{N}L}\mathbf{a}
\label{pclose}
\end{equation}%
in the basis $\mathbf{a}%
=(a_{r,11},...,a_{r,N_{y}N_{x}};a_{u,11},...,a_{u,N_{y}N_{x}};a_{l,11},...,a_{l,N_{y}N_{x}};a_{d,11},...,a_{d,N_{y}N_{x}})^{T}
$.

In the main text, we numerically solve Eqs. (\ref{cclose}) and (\ref{pclose}%
) for the linear closed network, i.e., $\chi =0$, and show the spectra $%
\mathcal{E}$ of the network with different geometries in Fig. 2 and 3. For
the closed nonlinear network, i.e., $\chi \neq 0$, the solutions are
unstable in general. In order to generate and stablize the state of light
with Kerr nonlinearities, we drive the network through the top boundary
mirrors of cylindrical open network.

\subsection{SMIIIB-Open network}

\begin{figure}[tbp]
\includegraphics[bb=18 474 581 725, width=10 cm, clip]{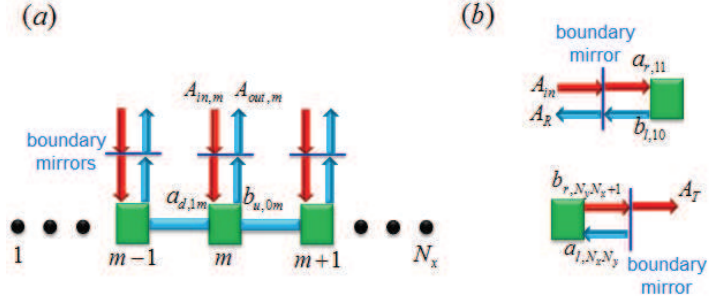}
\caption{(Color Online) The schematic for the boundary mirrors in the
cylindrical and planar networks: (a) Driven cylindrical network through each
of the partially transmissive mirrors on the top boundary. (b) Light
reflection and transmission through the boundary mirrors next to the nodes $%
(1,1)$ and $(N_{y},N_{x})$ in the planar network (i.e., nodes in the upper
left and lower right corners of the planar network).}
\label{boundary}
\end{figure}

For the open network on the cylinder shown in Fig. \ref{boundary}a, we drive
the network through the top boundary mirrors of the reflection
(transmission) coefficient $r_{\mathrm{BM}}$ ($t_{\mathrm{BM}}$), where the
driving light with frequency $\omega _{\mathrm{d}}$ has the amplitude $A_{%
\mathrm{in},m}^{(0)}$.

The $S$-matrix%
\begin{equation}
S_{\mathrm{BM}}=\left(
\begin{array}{cc}
t_{\mathrm{BM}} & ir_{\mathrm{BM}} \\
ir_{\mathrm{BM}} & t_{\mathrm{BM}}%
\end{array}%
\right)  \label{SBM}
\end{equation}%
of transmissive mirrors results in the boundary condition%
\begin{equation}
S_{\mathrm{BM}}\left(
\begin{array}{c}
A_{\mathrm{in},m}^{(0)} \\
b_{u,0m}e^{i\omega _{\mathrm{d}}L/2}%
\end{array}%
\right) =\left(
\begin{array}{c}
a_{d,1m}e^{-i\omega _{\mathrm{d}}L/2} \\
A_{\mathrm{out},m}^{(0)}%
\end{array}%
\right) ,
\end{equation}%
where $A_{\mathrm{in},m}^{(0)}$\ and $A_{\mathrm{out},m}^{(0)}$ are the
amplitudes of input and output fields above boundary mirrors (Fig. \ref%
{boundary}a), and $a_{d,1m}$ and $b_{u,0m}$ are the amplitudes of the
down-moving input field and the up-moving output field at the top of the
cylinder (i.e., the boundary nodes $(1,m)$).\ For the driving amplitude $A_{%
\mathrm{in},m}^{(0)}=A_{\mathrm{in}}^{(0)}e^{ik_{x}m}/\sqrt{N_{x}}$ with the
quasi-momentum $k_{x}$, the boundary condition becomes%
\begin{equation}
S_{\mathrm{BM}}\left(
\begin{array}{c}
A_{\mathrm{in}}^{(0)} \\
b_{u,0}e^{i\omega _{\mathrm{d}}L/2}%
\end{array}%
\right) =\left(
\begin{array}{c}
a_{d,1}e^{-i\omega _{\mathrm{d}}L/2} \\
A_{\mathrm{out}}^{(0)}%
\end{array}%
\right) ,  \label{SB}
\end{equation}%
where we used $a_{d,1m}=a_{d,1}e^{ik_{x}m}/\sqrt{N_{x}}$, $%
b_{u,0m}=b_{u,0}e^{ik_{x}m}/\sqrt{N_{x}}$, and $A_{\mathrm{out},m}^{(0)}=A_{%
\mathrm{out}}^{(0)}e^{ik_{x}m}/\sqrt{N_{x}}$. The scattering equation of the
open cylindrical network reads%
\begin{equation}
R_{\mathrm{BM}}S_{0}(k_{x})\mathbf{a}=e^{-i\omega _{\mathrm{d}}L}e^{i\frac{%
\chi }{N_{x}}\mathcal{N}_{n}L}\mathbf{a}-t_{\mathrm{BM}}e^{-i\omega _{%
\mathrm{d}}L/2}\mathbf{A}_{\mathrm{in}}^{(0)},  \label{copen}
\end{equation}%
where $R_{\mathrm{BM}}$ is obtained by replacing the diagonal matrix element
$I_{3N_{y}+1,3N_{y}+1}$ of the $4N_{y}$-dimensional identity matrix $I$ by $%
ir_{\mathrm{BM}}$, and $\mathbf{A}_{\mathrm{in}}^{(0)}=A_{\mathrm{in}}^{(0)}(%
\mathbf{0};\mathbf{0};\mathbf{0};\mathbf{1})^{T}$ is composed of the null
vector $\mathbf{0}$ and $\mathbf{1}=(1,0,...,0)$, which are of dimension $%
N_{y}$. The solution of the scattering Eq. (\ref{copen}) determines the
out-going amplitude%
\begin{equation}
A_{\mathrm{out}}^{(0)}=\frac{t_{\mathrm{BM}}}{ir_{\mathrm{BM}}}e^{-i\omega _{%
\mathrm{d}}L/2}a_{d,1}-\frac{A_{\mathrm{in}}^{(0)}}{ir_{\mathrm{BM}}}.
\end{equation}

When the driving frequency $\omega _{\mathrm{d}}$ is resonant with the
eigenfrequency $\mathcal{E}$ of the closed system, the boundary condition (%
\ref{SB}) leads to the amplitude%
\begin{equation}
a_{d,1}=\frac{t_{\mathrm{BM}}}{1-ir_{\mathrm{BM}}}A_{\mathrm{in}%
}^{(0)}e^{i\omega _{\mathrm{d}}L/2},
\end{equation}%
and the input-output relation $A_{\mathrm{out}}^{(0)}=e^{i\delta _{0}}A_{%
\mathrm{in}}^{(0)}$ with the phase shift%
\begin{equation}
\delta _{0}=\text{arg}(\frac{1+ir_{\mathrm{BM}}}{1-ir_{\mathrm{BM}}}).
\end{equation}

In the main text, we consider linear and nonlinear open networks on the
cylindrical geometry. In the linear case, we study the detection of the
energy spectrum through the phase shift $\delta _{0}$. In the nonlinear
case, we numerically solve Eq. (\ref{copen}) for the network with size $%
24\times 12$, and show the light distributions for different $k_{x}$ and $%
\omega _{\mathrm{d}}$ in Fig. 4.

For the open network on the plane shown in Fig. \ref{boundary}b, we drive
the network through the boundary mirror next to the node $(1,1)$ by the
light with frequency $\omega _{\mathrm{d}}$, and detect the transimission to
the node $(N_{x},N_{y})$.

The $S$-matrix (\ref{SBM}) of two transmissive mirrors (see Fig. \ref%
{boundary}b) gives the boundary conditions%
\begin{eqnarray}
S_{\mathrm{BM}}\left(
\begin{array}{c}
A_{\mathrm{in}}^{(0)} \\
b_{l,10}e^{i\omega _{\mathrm{d}}L/2}%
\end{array}%
\right) &=&\left(
\begin{array}{c}
a_{r,11}e^{-i\omega _{\mathrm{d}}L/2} \\
A_{\mathrm{R}}%
\end{array}%
\right) ,  \notag \\
S_{\mathrm{BM}}\left(
\begin{array}{c}
b_{r,N_{y}N_{x}+1}e^{i\omega _{\mathrm{d}}L/2} \\
0%
\end{array}%
\right) &=&\left(
\begin{array}{c}
A_{\mathrm{T}} \\
a_{l,N_{y}N_{x}}e^{-i\omega _{\mathrm{d}}L/2}%
\end{array}%
\right) ,
\end{eqnarray}%
where $A_{\mathrm{in}}^{(0)}$ denotes the input amplitude of the network,
and $A_{\mathrm{R}}$ ($A_{\mathrm{T}}$) are the reflection (transmission)
amplitude. The amplitudes of right-moving input and left-moving output
fields at the node $(1,1)$ are $a_{r,11}$ and $b_{l,10}$, while the
amplitudes of left-moving input and right-moving output fields at nodes $%
(N_{x},N_{y})$\ are $a_{l,N_{y}N_{x}}$\ and $b_{r,N_{y}N_{x}+1}$.\ The
scattering equation of the open network on the plane reads%
\begin{equation}
\bar{R}_{\mathrm{BM}}S_{0}\mathbf{a}=e^{-i\omega _{\mathrm{d}}L}e^{i\chi
\mathcal{N}L}\mathbf{a}-t_{\mathrm{BM}}e_{\mathrm{in}}^{-i\omega _{\mathrm{d}%
}L/2}\mathbf{A}_{\mathrm{in}}^{(0)},  \label{popen}
\end{equation}%
where $\bar{R}_{\mathrm{BM}}$ is obtained by replacing the diagonal matrix
elements $I_{1,1}$ and $I_{3N_{x}N_{y},3N_{x}N_{y}}$ of the $4N_{x}N_{y}$%
-dimensional identity matrix $I$ by $ir_{\mathrm{BM}}$, and $\mathbf{A}_{%
\mathrm{in}}^{(0)}=A_{\mathrm{in}}^{(0)}(\mathbf{1};\mathbf{0};\mathbf{0};%
\mathbf{0})^{T}$ is composed of the null vector $\mathbf{0}$ and $\mathbf{1}%
=(1,0,...,0)$, which are of dimension $N_{x}N_{y}$. The solution of the
scattering Eq. (\ref{popen}) determines the reflection and transmission
amplitudes%
\begin{eqnarray}
A_{\mathrm{R}} &=&\frac{t_{\mathrm{BM}}}{ir_{\mathrm{BM}}}%
a_{r,11}e^{-i\omega _{\mathrm{d}}L/2}-\frac{A_{\mathrm{in}}^{(0)}}{ir_{%
\mathrm{BM}}},  \notag \\
A_{\mathrm{T}} &=&\frac{t_{\mathrm{BM}}}{ir_{\mathrm{BM}}}%
a_{l,N_{y}N_{x}}e^{-i\omega _{\mathrm{d}}L/2}.
\end{eqnarray}

In the main text, we study the light transmission to the linear network on
the open plane by solving Eq. (\ref{popen}). The solution of Eq. (\ref{popen}%
) determines the light distribution in the network, and $\left\vert A_{%
\mathrm{T}}/A_{\mathrm{in}}^{(0)}\right\vert ^{2}$ is the transmission
probability, which are shown in Fig. 3b for different driving frequency $%
\omega _{\mathrm{d}}$.

\section{SMIV-Bogoliubov excitations}

In this section, we study the response of the cylindrical network to small
input fluctuations, where the total input field%
\begin{equation}
A_{m}^{\mathrm{in}}(t)=[A_{\mathrm{in},m}^{(0)}+\delta A_{m}^{\mathrm{in}%
}(t)]e^{-i\omega _{\mathrm{d}}t}
\end{equation}%
contains a small probe field%
\begin{equation}
\delta A_{m}^{\mathrm{in}}(t)=\frac{1}{\sqrt{N_{x}}}[\delta A_{\mathrm{in}%
}^{(+)}e^{ip_{x}m}e^{-i\omega _{\mathrm{f}}t}+\delta A_{\mathrm{in}%
}^{(-)}e^{i(2k_{x}-p_{x})m}e^{i\omega _{\mathrm{f}}t}]  \label{pf}
\end{equation}%
with frequency $\omega _{\mathrm{f}}$ around the driving field $A_{m}^{%
\mathrm{in}}$ that generates the steady state.

As discussed in Sec. SMIIA, the motion equations%
\begin{equation}
i\partial _{t}\delta \Psi _{H}+\mathbf{\Sigma }\partial _{x}\delta \Psi _{H}=%
\mathbf{M}_{H}\delta \Psi _{H},
\end{equation}%
and%
\begin{equation}
i\partial _{t}\delta \Psi _{V}+\mathbf{\Sigma }\partial _{x}\delta \Psi _{V}=%
\mathbf{M}_{V}\delta \Psi _{V}
\end{equation}%
describe the dynamics of Bogoliubov fluctuations $\delta \Psi _{H}=(\delta
\phi _{r,nm},\delta \phi _{l,nm-1},\delta \phi _{r,nm}^{\ast },\delta \phi
_{l,nm-1}^{\ast })^{T}$ and $\delta \Psi _{V}=(\delta \phi _{u,nm},\delta
\phi _{d,n+1m},\delta \phi _{u,nm}^{\ast },\delta \phi _{d,n+1m}^{\ast
})^{T} $ in the horizontal and vertical fibers, respectively. The probe
field (\ref{pf}) induces the fluctuation field with the configurations%
\begin{equation}
\delta \phi _{s,nm}(x,t)=\frac{1}{\sqrt{N_{x}}}[\delta \psi
_{s,n,p_{x}}(x)e^{ip_{x}m}e^{-i\omega _{\mathrm{f}}t}+\delta \psi
_{s,n,q_{x}}(x)e^{iq_{x}m}e^{i\omega _{\mathrm{f}}t}],
\end{equation}%
where $q_{x}=2k_{x}-p_{x}$ and the subindex $s=r,u,l,d$. The relations (\ref%
{Hf}) and (\ref{Vf}) become%
\begin{equation}
\bar{P}_{H}\left(
\begin{array}{c}
e^{-in\sigma \theta _{0}}\delta b_{r,n,p_{x}} \\
e^{-in\sigma \theta _{0}}\delta a_{l,n,p_{x}} \\
e^{in\sigma \theta _{0}}\delta b_{r,n,q_{x}}^{\ast } \\
e^{in\sigma \theta _{0}}\delta a_{l,n,q_{x}}^{\ast }%
\end{array}%
\right) =\left(
\begin{array}{c}
\delta a_{r,n,p_{x}} \\
\delta b_{l,n,p_{x}} \\
\delta a_{r,n,q_{x}}^{\ast } \\
\delta b_{l,n,q_{x}}^{\ast }%
\end{array}%
\right) ,  \label{Ph}
\end{equation}%
and%
\begin{equation}
\bar{P}_{V}\left(
\begin{array}{c}
\delta b_{u,n,p_{x}} \\
\delta a_{d,n+1,p_{x}} \\
\delta b_{u,n,q_{x}}^{\ast } \\
\delta a_{d,n+1,q_{x}}^{\ast }%
\end{array}%
\right) =\left(
\begin{array}{c}
\delta a_{u,n,p_{x}} \\
\delta b_{d,n+1,p_{x}} \\
\delta a_{u,n,q_{x}}^{\ast } \\
\delta b_{d,n+1,q_{x}}^{\ast }%
\end{array}%
\right) ,  \label{Pv}
\end{equation}%
where the input and output fields at the nodes are%
\begin{eqnarray}
\delta a_{r,n,p_{x}} &=&\delta \psi _{r,n,p_{x}}(L),\delta
a_{l,n,p_{x}}=e^{in\sigma \theta _{0}}\delta \psi _{l,n,p_{x}}(0),  \notag \\
\delta b_{r,n,p_{x}} &=&e^{in\sigma \theta _{0}}\delta \psi
_{r,n,p_{x}}(0),\delta b_{l,n,p_{x}}=\delta \psi _{l,n,p_{x}}(L),  \notag \\
\delta a_{u,n,p_{x}} &=&\delta \psi _{u,n,p_{x}}(L),\delta
a_{d,n+1,p_{x}}=\delta \psi _{d,n+1,p_{x}}(0),  \notag \\
\delta b_{u,n,p_{x}} &=&\delta \psi _{u,n,p_{x}}(0),\delta
b_{d,n+1,p_{x}}=\delta \psi _{d,n+1,p_{x}}(L).
\end{eqnarray}%
The propagation matrices $\bar{P}_{H}=e^{\mathbf{\Sigma }(\omega _{\mathrm{f}%
}-\mathbf{\bar{M}}_{H})L}$ and $\bar{P}_{V}=e^{\mathbf{\Sigma }(\omega _{%
\mathrm{f}}-\mathbf{\bar{M}}_{V})L}$ are determined by
\begin{equation}
\mathbf{\bar{M}}_{H}=\frac{\chi }{N_{x}}\left(
\begin{array}{cccc}
\left\vert a_{r,n}\right\vert ^{2} & 2e^{i(k_{x}-p_{x})}\tilde{a}%
_{l,n}^{\ast }a_{r,n} & a_{r,n}^{2} & 2e^{i(k_{x}-p_{x})}a_{r,n}\tilde{a}%
_{l,n} \\
2e^{-i(k_{x}-p_{x})}a_{r,n}^{\ast }\tilde{a}_{l,n} & \left\vert \tilde{a}%
_{l,n}\right\vert ^{2} & 2e^{-i(k_{x}-p_{x})}a_{r,n}\tilde{a}_{l,n} & \tilde{%
a}_{l,n}^{2} \\
-a_{r,n}^{\ast 2} & -2e^{i(k_{x}-p_{x})}a_{r,n}^{\ast }\tilde{a}_{l,n}^{\ast
} & -\left\vert a_{r,n}\right\vert ^{2} & -2e^{i(k_{x}-p_{x})}a_{r,n}^{\ast }%
\tilde{a}_{l,n} \\
-2e^{-i(k_{x}-p_{x})}a_{r,n}^{\ast }\tilde{a}_{l,n}^{\ast } & -\tilde{a}%
_{l,n}^{\ast 2} & -2e^{-i(k_{x}-p_{x})}\tilde{a}_{l,n}^{\ast }a_{r,n} &
-\left\vert \tilde{a}_{l,n}\right\vert ^{2}%
\end{array}%
\right) ,
\end{equation}%
and%
\begin{equation}
\mathbf{\bar{M}}_{V}=\frac{\chi }{N_{x}}\left(
\begin{array}{cccc}
\left\vert a_{u,n}\right\vert ^{2} & 2a_{d,n+1}^{\ast }a_{u,n} & a_{u,n}^{2}
& 2a_{u,n}a_{d,n+1} \\
2a_{u,n}^{\ast }a_{d,n+1} & \left\vert a_{d,n+1}\right\vert ^{2} &
2a_{u,n}a_{d,n+1} & a_{d,n+1}^{2} \\
-a_{u,n}^{\ast 2} & -2a_{u,n}^{\ast }a_{d,n+1}^{\ast } & -\left\vert
a_{u,n}\right\vert ^{2} & -2a_{u,n}^{\ast }a_{d,n+1} \\
-2a_{u,n}^{\ast }a_{d,n+1}^{\ast } & -a_{d,n+1}^{\ast 2} & -2a_{d,n+1}^{\ast
}a_{u,n} & -\left\vert a_{d,n+1}\right\vert ^{2}%
\end{array}%
\right)
\end{equation}%
in the horizontal and vertical fibers, respectively, where $\tilde{a}%
_{l,n}=e^{-in\sigma \theta _{0}}a_{l,n}$.

The relations between the input $\delta a$ and output fields $\delta b$ at
the nodes follow from Eqs. (\ref{Ph}) and (\ref{Pv}) as%
\begin{equation}
\left(
\begin{array}{cccc}
\bar{P}_{H,11}e^{-in\sigma \theta _{0}} & 0 & \bar{P}_{H,13}e^{in\sigma
\theta _{0}} & 0 \\
-\bar{P}_{H,21}e^{-in\sigma \theta _{0}} & 1 & -\bar{P}_{H,23}e^{in\sigma
\theta _{0}} & 0 \\
\bar{P}_{H,31}e^{-in\sigma \theta _{0}} & 0 & \bar{P}_{H,33}e^{in\sigma
\theta _{0}} & 0 \\
-\bar{P}_{H,41}e^{-in\sigma \theta _{0}} & 0 & -\bar{P}_{H,43}e^{in\sigma
\theta _{0}} & 1%
\end{array}%
\right) \left(
\begin{array}{c}
\delta b_{r,n,p_{x}} \\
\delta b_{l,n,p_{x}} \\
\delta b_{r,n,q_{x}}^{\ast } \\
\delta b_{l,n,q_{x}}^{\ast }%
\end{array}%
\right) =\left(
\begin{array}{cccc}
1 & -\bar{P}_{H,12}e^{-in\sigma \theta _{0}} & 0 & -\bar{P}%
_{H,14}e^{in\sigma \theta _{0}} \\
0 & \bar{P}_{H,22}e^{-in\sigma \theta _{0}} & 0 & \bar{P}_{H,24}e^{in\sigma
\theta _{0}} \\
0 & -\bar{P}_{H,32}e^{-in\sigma \theta _{0}} & 1 & -\bar{P}%
_{H,34}e^{in\sigma \theta _{0}} \\
0 & \bar{P}_{H,42}e^{-in\sigma \theta _{0}} & 0 & \bar{P}_{H,44}e^{in\sigma
\theta _{0}}%
\end{array}%
\right) \left(
\begin{array}{c}
\delta a_{r,n,p_{x}} \\
\delta a_{l,n,p_{x}} \\
\delta a_{r,n,q_{x}}^{\ast } \\
\delta a_{l,n,q_{x}}^{\ast }%
\end{array}%
\right) ,  \label{Hfc}
\end{equation}%
and%
\begin{equation}
\left(
\begin{array}{cccc}
\bar{P}_{V,11} & 0 & \bar{P}_{V,13} & 0 \\
-\bar{P}_{V,21} & 1 & -\bar{P}_{V,23} & 0 \\
\bar{P}_{V,31} & 0 & \bar{P}_{V,33} & 0 \\
-\bar{P}_{V,41} & 0 & -\bar{P}_{V,43} & 1%
\end{array}%
\right) \left(
\begin{array}{c}
\delta b_{u,n,p_{x}} \\
\delta b_{d,n+1,p_{x}} \\
\delta b_{u,n,q_{x}}^{\ast } \\
\delta b_{d,n+1,q_{x}}^{\ast }%
\end{array}%
\right) =\left(
\begin{array}{cccc}
1 & -\bar{P}_{V,12} & 0 & -\bar{P}_{V,14} \\
0 & \bar{P}_{V,22} & 0 & \bar{P}_{V,24} \\
0 & -\bar{P}_{V,32} & 1 & -\bar{P}_{V,34} \\
0 & \bar{P}_{V,42} & 0 & \bar{P}_{V,44}%
\end{array}%
\right) \left(
\begin{array}{c}
\delta a_{u,n,p_{x}} \\
\delta a_{d,n+1,p_{x}} \\
\delta a_{u,n,q_{x}}^{\ast } \\
\delta a_{d,n+1,q_{x}}^{\ast }%
\end{array}%
\right) .  \label{Vfc}
\end{equation}%
Equations (\ref{Hfc}) and (\ref{Vfc}) give the relation%
\begin{equation}
\left(
\begin{array}{c}
\delta b \\
\delta b^{\ast }%
\end{array}%
\right) =H_{\mathrm{f}}(\omega _{\mathrm{f}})\left(
\begin{array}{c}
\delta a \\
\delta a^{\ast }%
\end{array}%
\right)  \label{HF}
\end{equation}

On the other hand, the $S$-matrix $S_{\mathrm{node}}$ leads to the
input-output relation%
\begin{equation}
S_{f,\mathrm{node}}(p_{x})\left(
\begin{array}{c}
\delta a_{r,n,p_{x}} \\
\delta a_{u,n,p_{x}} \\
\delta a_{l,n,p_{x}} \\
\delta a_{d,n,p_{x}}%
\end{array}%
\right) =\left(
\begin{array}{c}
e^{ip_{x}}\delta b_{r,n,p_{x}} \\
\delta b_{u,n-1,p_{x}} \\
e^{-ip_{x}}\delta b_{l,n,p_{x}} \\
\delta b_{d,n+1,p_{x}}%
\end{array}%
\right)  \label{Snf}
\end{equation}%
at each node, where%
\begin{equation}
S_{f,\mathrm{node}}(p_{x})=e^{i\omega _{\mathrm{d}}L}e^{-i\frac{\chi }{N_{x}}%
\mathcal{N}_{n}L}\left(
\begin{array}{cccc}
e^{-ip_{x}} & 0 & 0 & 0 \\
0 & 1 & 0 & 0 \\
0 & 0 & e^{ip_{x}} & 0 \\
0 & 0 & 0 & 1%
\end{array}%
\right) S_{\mathrm{node}}.
\end{equation}%
The $S$-matrix $S_{\mathrm{BM}}$ of transmissive mirrors at the top of the
cylinder give the boundary condition%
\begin{eqnarray}
ir_{\mathrm{BM}}\delta b_{u,0,p_{x}} &=&e^{-i\omega _{\mathrm{f}}L}\delta
a_{d,1,p_{x}}-t_{\mathrm{BM}}\delta A_{\mathrm{in}}^{(+)}e^{i(\omega _{%
\mathrm{d}}-\omega _{\mathrm{f}})\frac{L}{2}},  \notag \\
ir_{\mathrm{BM}}\delta b_{u,0,q_{x}} &=&e^{i\omega _{\mathrm{f}}L}\delta
a_{d,1,q_{x}}-t_{\mathrm{BM}}\delta A_{\mathrm{in}}^{(-)}e^{i(\omega _{%
\mathrm{d}}+\omega _{\mathrm{f}})\frac{L}{2}}.  \label{top}
\end{eqnarray}%
and the components%
\begin{eqnarray}
\delta A_{\mathrm{out}}^{(+)} &=&\frac{t_{\mathrm{BM}}}{ir_{\mathrm{BM}}}%
e^{-i(\omega _{\mathrm{d}}+\omega _{\mathrm{f}})\frac{L}{2}}\delta
a_{d,1,p_{x}}-\frac{\delta A_{\mathrm{in}}^{(+)}}{ir_{\mathrm{BM}}},  \notag
\\
\delta A_{\mathrm{out}}^{(-)} &=&\frac{t_{\mathrm{BM}}}{ir_{\mathrm{BM}}}%
e^{-i(\omega _{\mathrm{d}}-\omega _{\mathrm{f}})\frac{L}{2}}\delta
a_{d,1,q_{x}}-\frac{\delta A_{\mathrm{in}}^{(-)}}{ir_{\mathrm{BM}}}.
\label{Aout}
\end{eqnarray}%
of positive and negative frequencies in the output fluctuation field%
\begin{equation}
\delta A_{m}^{\mathrm{out}}(t)=\frac{1}{\sqrt{N_{x}}}[\delta A_{\mathrm{out}%
}^{(+)}e^{ip_{x}m}e^{-i\omega _{\mathrm{f}}t}+\delta A_{\mathrm{out}%
}^{(-)}e^{iq_{x}m}e^{i\omega _{\mathrm{f}}t}]
\end{equation}%
around the output field $A_{m}^{\mathrm{out}}$ in the steady state.

\begin{figure}[tbp]
\includegraphics[bb=18 493 555 759, width=10 cm, clip]{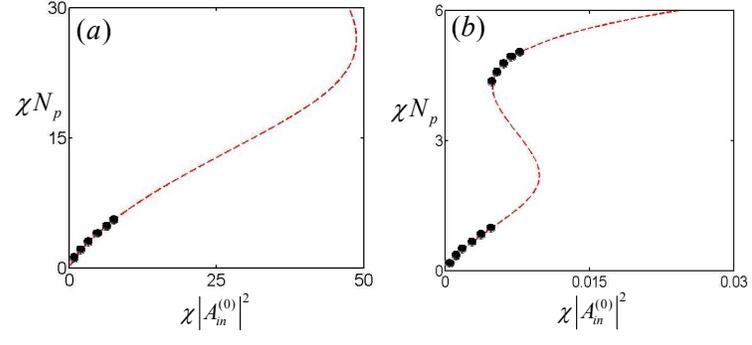}
\caption{(Color Online) For $k_{x}=0.26$ and $r_{\mathrm{BM}}=0.9$, the
stable regimes are shown by the black dots in the $\protect\chi N_{p}$
versus $\protect\chi \left\vert A_{\mathrm{in}}^{(0)}\right\vert ^{2}$
curves, where $\protect\omega _{\mathrm{d}}=0.22$ (a) and $\protect\omega _{%
\mathrm{d}}=4.5\times 10^{-2}$ (b).}
\label{stable}
\end{figure}

By the elimination of the output fluctuation fields $\delta
b_{s,n,p_{x}(q_{x})}$ in Eqs. (\ref{HF}), (\ref{Snf}), and the boundary
condition (\ref{top}), we establish the scattering equation%
\begin{equation}
\mathbf{D}(\omega _{\mathrm{f}})\left(
\begin{array}{c}
\delta \mathbf{a}_{p_{x}} \\
\delta \mathbf{a}_{q_{x}}^{\ast }%
\end{array}%
\right) =t_{\mathrm{BM}}\left(
\begin{array}{c}
\delta \mathbf{A}_{\mathrm{in}}^{(+)}e^{i(\omega _{\mathrm{d}}-\omega _{%
\mathrm{f}})\frac{L}{2}} \\
\delta \mathbf{A}_{\mathrm{in}}^{(-)\ast }e^{-i(\omega _{\mathrm{d}}+\omega
_{\mathrm{f}})\frac{L}{2}}%
\end{array}%
\right)  \label{DD}
\end{equation}%
of Bogoliubov fluctuations by the matrix%
\begin{equation}
\mathbf{D}(\omega _{\mathrm{f}})=H_{\mathrm{f}}(\omega _{\mathrm{f}})-\left(
\begin{array}{cc}
S_{f,\mathrm{node}}(p_{x}) & 0 \\
0 & S_{f,\mathrm{node}}^{\ast }(q_{x})%
\end{array}%
\right) ,
\end{equation}%
where $\delta \mathbf{a}_{p_{x}}=(\delta a_{r,n,p_{x}},\delta
a_{u,n,p_{x}},\delta a_{l,n,p_{x}},\delta a_{d,n,p_{x}})^{T}$, and $\delta
\mathbf{A}_{\mathrm{in}}^{(\pm )}=\delta A_{\mathrm{in}}^{(\pm )}(\mathbf{0};%
\mathbf{0};\mathbf{0};\mathbf{1})$ is composed of the null vector $\mathbf{0}
$ and $\mathbf{1}=(1,0,...,0)$, which are of dimension $N_{y}$. The
steady-state is stable if all roots $\mathcal{E}_{f}$ of det$\mathbf{D}%
(\omega _{\mathrm{f}})$ satisfy Im$\mathcal{E}_{f}<0$. For $\omega _{\mathrm{%
d}}=0.22$ and $\omega _{\mathrm{d}}=4.5\times 10^{-2}$, we mark the stable
regimes by the black dots in the $\chi N_{p}$ versus $\chi \left\vert A_{%
\mathrm{in}}^{(0)}\right\vert ^{2}$ curves of Fig. \ref{stable}, where $%
k_{x}=0.26$ and $r_{\mathrm{BM}}=0.9$.

The solutions $(\delta \mathbf{a}_{p_{x}},\delta \mathbf{a}_{q_{x}}^{\ast })$
of Eq. (\ref{DD}) lead to the input-output relation%
\begin{equation}
\left(
\begin{array}{c}
\delta A_{\mathrm{out}}^{(+)} \\
\delta A_{\mathrm{out}}^{(-)\ast }%
\end{array}%
\right) =M_{\mathrm{IO}}\left(
\begin{array}{c}
\delta A_{\mathrm{in}}^{(+)} \\
\delta A_{\mathrm{in}}^{(-)\ast }%
\end{array}%
\right)
\end{equation}%
by Eqs. (\ref{Aout}), where the $2\times 2$ matrix%
\begin{equation}
M_{\mathrm{IO}}=\frac{1}{ir_{\mathrm{BM}}}\sigma _{z}[t_{\mathrm{BM}%
}^{2}e^{-i\omega _{\mathrm{f}}L}e^{-i\omega _{\mathrm{d}}\frac{L}{2}\sigma
_{z}}\tilde{D}(\omega )e^{i\omega _{\mathrm{d}}\frac{L}{2}\sigma _{z}}-I_{2}]
\end{equation}%
is determined by%
\begin{equation}
\tilde{D}(\omega )=\left(
\begin{array}{cc}
\lbrack \mathbf{D}^{-1}(\omega )]_{3N_{y}+1,3N_{y}+1} & [\mathbf{D}%
^{-1}(\omega )]_{3N_{y}+1,7N_{y}+1} \\
\lbrack \mathbf{D}^{-1}(\omega )]_{7N_{y}+1,3N_{y}+1} & [\mathbf{D}%
^{-1}(\omega )]_{7N_{y}+1,7N_{y}+1}%
\end{array}%
\right) .
\end{equation}

By using the input-output matrix $M_{\mathrm{IO}}$, we investigate the
reflection fields with the positive and negative frequencies under the
driving positive frequency field ($\delta A_{\mathrm{in}}^{(-)}=0$). The
components $(M_{\mathrm{IO}})_{11}$ and $(M_{\mathrm{IO}})_{21}$ determine
the squeezing spectra shown in Fig. 5a of the main text. By solving Eq. (\ref%
{DD}), in Fig. 5b of the main text, we show the light distributions of
Bogoliubov fluctuations around steady states.


\begin{thebibliography}{99}
\bibitem{QT} We emphasize that our results from the fluctuation analysis for
the classical light are also valid in the quantum regime, where the
squeezing behaviors of quantum edge fluctuations arise from the interplay of
the Kerr nonlinearities and topological effects.
\end{thebibliography}
\end{document}